\pdfoutput=1
\documentclass[aps,pra,twocolumn,superscriptaddress,a4paper,floatfix]{revtex4}
\usepackage[latin1]{inputenc}
\usepackage{graphicx}
\usepackage{amsmath,amssymb}
\usepackage{hyperref}

\newcommand \beq{\begin{eqnarray}}
\newcommand \eeq{\end{eqnarray}}

\begin{document}

\title{Universal Noise in Continuous Transport Measurements of Interacting Fermions}

\author{Shun Uchino}
\affiliation{Weseda Institute for Advanced Study, Waseda University, Shinjuku, Tokyo 169-8050, Japan}
\affiliation{RIKEN Center for Emergent Matter Science, Wako, Saitama 351-0198, Japan}
\author{Masahito Ueda}
\affiliation{RIKEN Center for Emergent Matter Science, Wako, Saitama 351-0198, Japan}
\affiliation{Department of Physics, University of Tokyo, Hongo, Bunkyo-ku, Tokyo
113-0033, Japan}
\author{Jean-Philippe Brantut}
\affiliation{Institute of Physics, EPFL, 1015 Lausanne, Switzerland}


\begin{abstract}
We propose and analyze continuous measurements of atom number and atomic currents using dispersive probing in an optical cavity. For an atom-number measurement in a closed system, we relate both the detection noise and the heating rate due to measurement back-action to Tan's contact, and identify an emergent universal quantum non-demolition (QND) regime in the good-cavity limit. We then show that such a continuous QND measurement of atom number serves as a quantum-limited current transducer in a two-terminal setup. We derive a universal bound on the precision of current measurement, which results from a tradeoff between detection noise and back-action of the atomic current measurement. Our results apply regardless of the strength of interaction  or the state of matter and set fundamental bounds on future precision measurements of transport properties in cold-atom quantum simulators. 

\end{abstract}


\maketitle

\section{Introduction}

Transport is among the best probes of quantum many-body systems, because it is sensitive to the nature of the system's excitations. Recently, quantum simulation has emerged as a new method to tackle the many-body problem by using a controlled cold atomic gas as a model for electrons in condensed matter \cite{Cirac:2012aa}. With this approach getting ready to address the open questions of condensed matter physics, there is a growing interest in the direct measurement of transport properties in atomic gases \cite{Chien:2015ab,0953-8984-29-34-343003}. The current methods used to investigate transport in cold atomic gases suffer from the intrinsically destructive nature of the observation. Even when snapshots of the density distribution are obtained at the level of individual atoms, the investigation of the dynamics involves sample-to-sample fluctuations. As a result, the noise in the preparation directly feeds in the measurement outcomes, rendering the cold-atom transport measurements far less precise than their condensed-matter counterparts~\cite{BLANTER20001}. 

\begin{figure}[htbp]
\begin{center}
\includegraphics[width=0.3\textwidth]{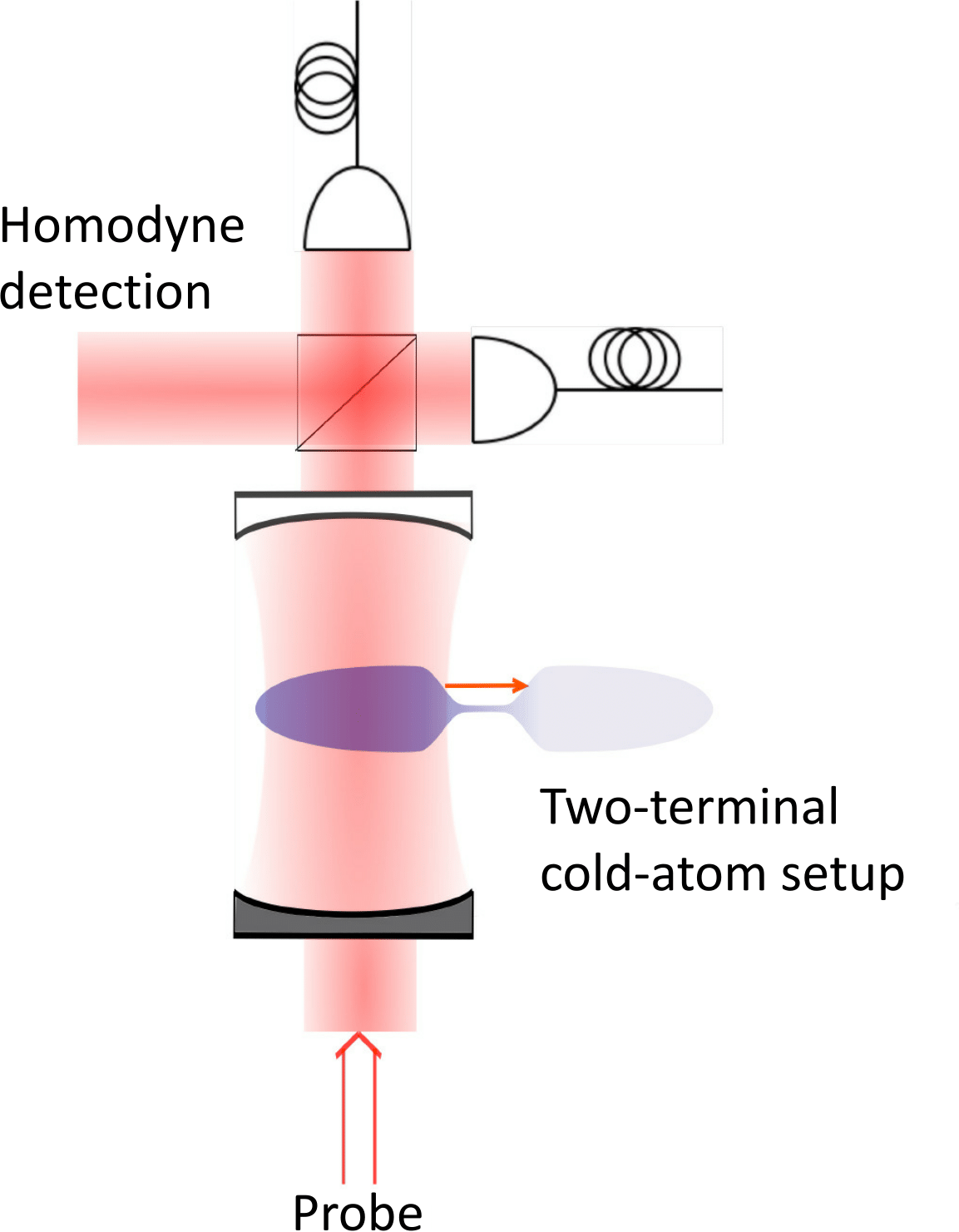}
\caption{Experimental concept of the atomic current detector. An atomic cloud is shaped in a two-terminal setup with large reservoirs connected by a mesoscopic channel. One reservoir overlaps with the mode of a single-sided optical cavity. A chemical potential difference introduced between the two reservoirs drives a quasi-DC current in the channel. 
A probe beam far detuned from the atomic resonance is sent onto the cavity, and its phase, measured by homodyne detection, provides a real-time measurement of the atom number in the reservoir.}
\label{fig:twoTerms}
\end{center}
\end{figure}

In this paper, we describe a method for the continuous measurement of atomic currents over single realizations of a quantum gas, that applies equally well to weakly and strongly interacting gases at the ultimate limit set by quantum-mechanical back-action. 
The concept is depicted in Fig. \ref{fig:twoTerms}: it relies on (i) the two (or multi-)terminal configuration, where the system of interest is connected to large atomic reservoirs allowing to inject and collect particles, and (ii) the use of continuous measurements of atom numbers using a high-finesse cavity and a probe laser far from the atomic resonance. The atomic current consists of atoms continuously entering and leaving the reservoir, thereby interacting with the cavity mode and causing the phase shift of a probe laser, which is measured by a quantum-limited interferometer. The high-finesse cavity ensures that the phase shift and the measurement back-action do not suffer 
significantly from the effects of spontaneous emission \cite{Lye:2003aa,2011AAMOP..60..201T}. 

Outline of this paper is as follows.

In Sec.~II, we investigate the continuous dispersive measurement of a reservoir in the absence of currents, and express both the noise spectrum and the heating due to measurement back-action in terms of Tan's contact--the parameter relating the two-body correlations at short distance to the macroscopic properties of the cloud \cite{Tan:2008ac,braaten2012universal}. The connection between Tan's contact and measurement back action provides a quantitative estimate of the measurement outcomes, and allows one to identify a good-cavity regime where the back-action vanishes, realizing a emergent QND measurement. Similar to Tan's relations, this applies to an arbitrary
strength of interaction, demonstrating the universal character of the measurement scheme. This complements existing proposals focusing on single-particle physics or lattice systems \cite{Eckert:2008aa,PhysRevA.87.021601,PhysRevA.90.023628,PhysRevLett.102.020403,PhysRevLett.115.060401,PhysRevLett.115.095301,Yang:2017aa}.


In Sec.~III, we consider the reservoirs connected by a channel which carries atomic currents. There, even in the good-cavity regime, the observation of a reservoir produces a back-action on the transport process, which we interpret as fluctuations of the chemical-potential bias across the channel. Together with the intrinsic detection noise, this back-action yields a finite, universal quantum limit on the precision of the atomic current measurement. Although the atom-field coupling is treated within linear response theory, this result assume neither Fermi-liquid behavior nor linear response of the atomic current to the applied bias. We express the limit explicitly in terms of a finite-bias admittance of the channel.

Section~IV discusses possible experiments with cold atomic gases.
In the appendices, we discuss technical details of our formulation.



\section{Continuous reservoir observation} We first consider a closed reservoir containing $N$ fermions at zero temperature, dispersively coupled to the optical field in a Fabry-Perot cavity which, in turn, is coupled to an environment through an imperfect mirror. The field is far detuned from the atomic resonance such that the excited states of the atoms are not populated. The atoms populate equally two hyperfine states, and for simplicity we assume these two states to be coupled identically to the field, though this assumption is not essential. Then the Hamiltonian of the system reads \cite{Ritsch:2013aa}
\begin{equation}
\hat{H} = \hat{H}_{at} +  \omega_c \hat{d}^\dagger \hat{d} +  \Omega \hat{M}  \hat{d}^\dagger \hat{d}.
\end{equation}
Here $\hat{H}_{at}$ is the Hamiltonian of the atoms in the absence of the cavity field, consisting of the kinetic energy, the interaction energy between the two spin components, and a possible trapping potential. The empty cavity has frequency $ \omega_c$, and $ \hat{d}$ ($ \hat{d}^\dagger$) annihilates (creates) a photon in the cavity. The coupling $\Omega$ of the atoms to the field represents the shift of the cavity resonance due to the presence of one atom maximally coupled to the field, and $\hat{M}$ is the overlap of the density distribution of the atoms with the cavity mode:
\begin{equation}
\hat{M} = \int d\mathbf{r} \cos^2(kz) \hat{n}(\mathbf{r}) =  \frac{1}{2}\hat{N} + \frac 1 4 (\hat{n}_{2k} + \hat{n}_{-2k}),
\end{equation}
where $k$ is the wave vector of cavity photons. We have introduced the operators for the total atom density $\hat{n}(\mathbf{r})$, the total atom number $\hat{N}$ and the density fluctuations $\hat{n}_{2k} = \int d\mathbf{r} e^{-2ikz} \hat{n}(\mathbf{r})$. Here we consider a situation in which the waist of the cavity mode is much larger than the atomic cloud.

We describe a coherent driving resonant with the cavity and the coupling to the vacuum using the input-output formalism \cite{gardiner2004}. We decompose the atom-field coupling into a non-fluctuating part which we include in $\hat{H}_{at}$~\footnote{This implies the presence of a static lattice modulation. For $k\gg k_F$ the modulation is negligible as long as the number of intra-cavity photons is small compared with $ \hbar k^2/2m\Omega$. Moreover, the static lattice can experimentally be suppressed using interrogation techniques described in Refs. \cite{PhysRevA.94.061601,1367-2630-19-8-083002}} and a part containing the vacuum fluctuations $ \hat{\eta}$. To first order in $\hat{\eta}$, the coupling Hamiltonian reads $ \hat{F} \hat{M}$, where  $\hat{F}=ig ( \hat{\eta}^\dagger - \hat{\eta}) $ with $g = 2 \Omega \sqrt{\Phi/\kappa}$ being the measurement strength, $\Phi$ is the photon flux incident on the cavity, and $\kappa$ is the cavity decay rate. Importantly, to first order in  fluctuations, the time evolution of $\hat{F}$ is decoupled from that of the atoms, so that the freely evolving $\hat{F} (t)$ can be directly treated as a perturbation for the atoms. 

The presence of the operators $\hat{n}_{\pm 2k}$ in $\hat{M}$ implies that $\hat{M}$ does not commute with the atomic Hamiltonian, and that the measurement is thus destructive. In fact, the Heisenberg equation directly relates the commutator of $\hat{M}$ with $\hat{H}_{at}$ to the energy absorption rate due to measurement: 
\begin{equation}
\frac{d \hat{H}_{at} }{dt} = - i  \left[ \hat{H}_{at}, \hat{M} \right ] \hat{F}.
\end{equation}
We evaluate this expression using linear response theory, obtaining (see  Appendix B for details)
\begin{equation}
\frac{d\mathcal{E}_{\mathrm{at}}}{dt}= -\frac{g^2\kappa}{16n}\chi^{R}(2k,i\kappa/2),
\end{equation}
where $\mathcal{E}_{\mathrm{at}}$ is the energy per atom, $n$ is the atomic density, and $\chi^R$ is the retarded density response function which is determined by an equilibrium average \footnote{It might seem that the heating rate given by the response at momentum $2k$ is due to the cosine shape of the mode function in the Fabry-Perot configuration.  In fact, heating is due to the effect of photon back-scattering onto atoms, imparting momentum $2k$ to the cloud, which arises regardless of the mode profile. Choosing instead a ring cavity yields the same result up to a numerical factor as discussed in Appendix E. }.

We now focus on the case with $2k\gg k_F$ and $\kappa/2\gg \epsilon_F$, which we expect to be realized in typical cold-atom experiments \cite{2007Natur.450..272C,Brennecke:2007aa,PhysRevA.75.063620,Murch:2008aa} (note however Ref.~\cite{Keler:2014aa}). In this regime, the density response function can systematically be evaluated using the operator product expansion~\cite{PhysRevLett.100.205301,PhysRevA.81.063634,PhysRevA.84.043603,PhysRevA.85.013613,PhysRevX.7.011022} (OPE), regardless of the interaction strength, temperature, and phase of matter. The result of this expansion is expressed as
\begin{multline}
\frac{1}{g^2}\frac{d\mathcal{E}_{\mathrm{at}}}{dt} =  g_n(x) + g_c(x,2k a)\frac{C}{k_F^4} \left( \frac{k_F}{2k}\right) \\
+g_H(x) \frac{\mathcal{E}_{\mathrm{at}}}{\mathcal{E}_{0\mathrm{at}}}\left( \frac{k_F}{2k}\right)^2+..., 
\label{eq:heating}
\end{multline}
where $C$ is Tan's contact density, $\mathcal{E}_{0\mathrm{at}} = 3\epsilon_F/5$ is the energy per atom of a noninteracting Fermi gas, $g_n$ is a function of $x=\frac{\kappa m}{4 k^2}$, independent of interactions, and $g_c$, and $g_H$ are universal functions of $x$ and $ka$, where $a$ is the $s$-wave scattering length.  The analytic expressions of these functions are shown in Appendix C. 

\begin{figure}[htbp]
\begin{center}
\includegraphics[width=0.45\textwidth]{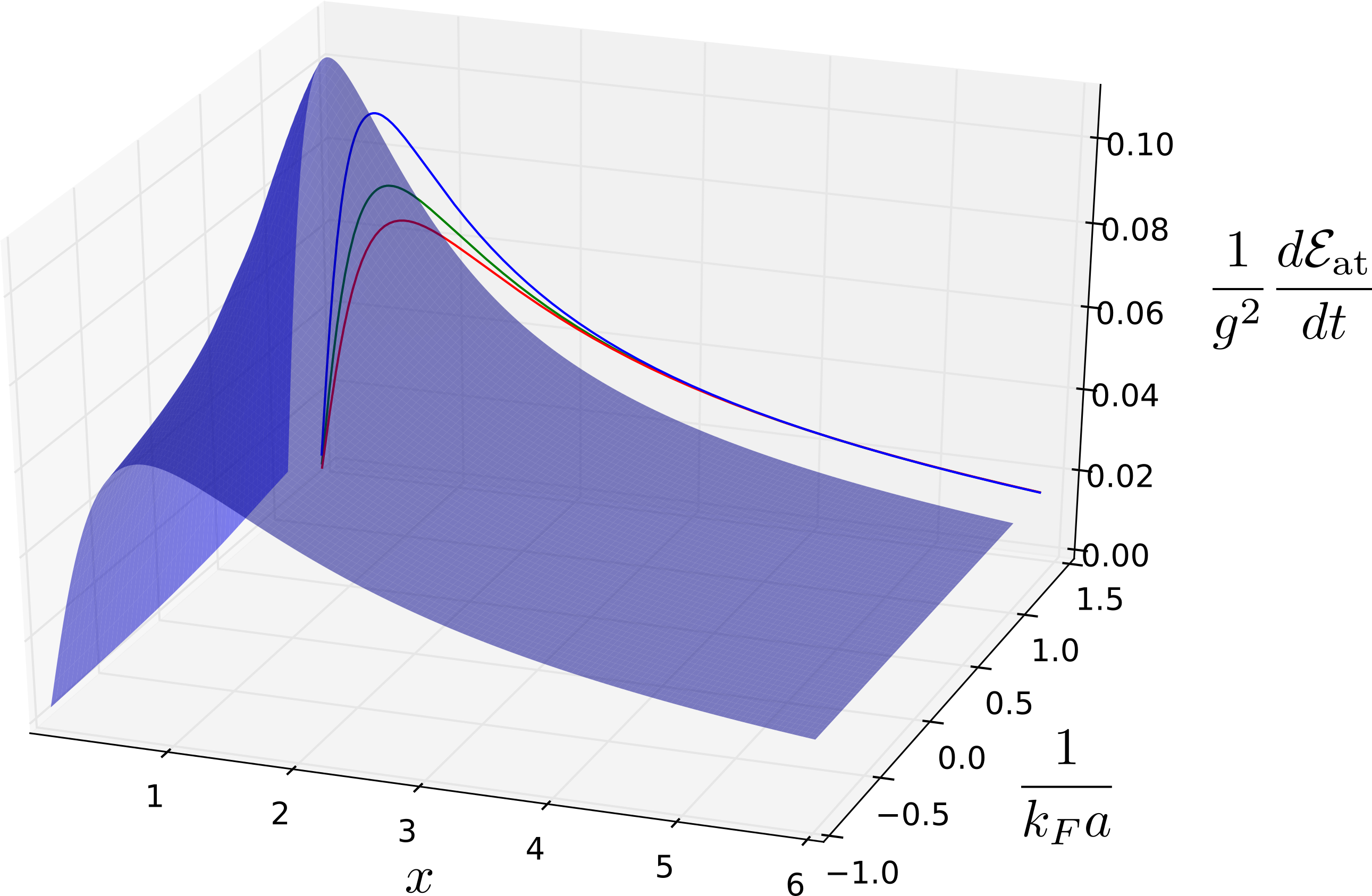}
\caption{ Dimensionless energy absorption rate as functions of $x$ and the interaction parameter $1/(k_Fa)$. The cuts in red, green and blue present the dimensionless energy absorption rate for $1/(k_Fa)=-0.5,\,0,$ and $0.5$, respectively.}
\label{fig:heating_rate}
\end{center}
\end{figure}

Figure \ref{fig:heating_rate} presents the right-hand side of Eq.~\eqref{eq:heating} for the ratio $k_F/(2k) = 0.2$, in line with typical experiments using $^6$Li atoms, and well in the valid regime of the OPE~\cite{PhysRevLett.109.050403,Hoinka:2013aa}. The normalization implies that variations of $x$ (i.e., $\kappa$) are taken at a constant mean intra-cavity photon number. In the regime of large $x$, the heating rate tends to vanish according to a power law, a feature expected from Tan's relations. In the opposite regime $x<1$, heating linearly decreases to zero due to the saturation of $\chi$ at low $\omega$. There, the cavity responds too slowly to resolve the atomic motion along the cavity mode, thereby realizing an emergent QND measurement \cite{PhysRevA.43.3819,PhysRevLett.120.133601}. 
Importantly, the conditions for both QND operation $\left(\kappa<\frac{4k^2}{m}\right)$ and the validity of the OPE $(\epsilon_F<\kappa)$ can be fulfilled simultaneously, since $\epsilon_F$ and $\frac{4k^2}{m}$ differ by more than one order of magnitude in typical experiments \cite{PhysRevLett.109.050403,Hoinka:2013aa}.


As interactions are varied from the BCS to the BEC limit, the maximum shifts towards low frequencies, as a result of pairing becoming more pronounced. Here, we have neglected heating due to spontaneous emission~\cite{Gerbier:2010aa}, which is justified for cavities with large enough cooperativities (see Appendix F for details).

The input-output formalism predicts that the photocurrent of the homodyne detector can be expressed after appropriate renormalization and up to a constant offset as $I_h(t) =  \hat{b}_{\mathrm{out}} + \hat{b}^\dagger_{\mathrm{out}}$. Here $\hat{b}_{\mathrm{out}}$ describes the field emanating from the cavity, and the phase of the interferometer is chosen to be zero.
The mean homodyne current relates to the atom number through $\langle \hat{N} \rangle = \sqrt{\kappa} \langle I_h \rangle/ 2g$. Using again linear response theory, we relate the photocurrent noise spectral density, referred back to the atom number, to the dynamical structure factor $S$ of the gas (see Appendix D for details):
\begin{multline}
\mathcal{S}_{NN}(\omega) =  \frac{\kappa}{4g^2} + N^2 \delta(\omega)+ \frac {V} 4 \frac{S(2k,\omega) + S(2k,-\omega)}{1 + \left(\frac{2\omega}{\kappa}\right)^2}, 
\label{eq:noise}
\end{multline}
where $V$ is the volume of the system. The first term on the right-hand side shows the imprecision introduced by the photon shot noise, the second term arises from the constant value of $N$, and the last term represents the quantum fluctuations of atoms in the cavity mode. While a similar form of the noise was obtained in a Bose-Einstein condensate inside a cavity from a different perspective~\cite{landig2015aa}, we do not rely on the mean-field approximation, and the above formula is valid for any system in the weak atom-field coupling regime. 


Similar to the heating rate, the atomic contribution to the noise is also universal in the regime where the OPE is valid. In contrast to the density response at imaginary frequencies, the OPE expansion for the structure factor has been considered in Refs.~\cite{PhysRevA.81.063634,PhysRevA.84.043603,PhysRevA.85.013613,PhysRevX.7.011022}. In the good cavity regime, the contribution of the atomic fluctuations becomes negligible, since $S$ decreases according to a power law in the low-frequency regime, confirming the emergent QND character of the measurement. Our analysis applies for any interactions between fermions in the weak-measurement limit. In the opposite, strong-measurement regime, even the non-interacting Fermi gas shows large nonlinearities \cite{PhysRevA.78.023815,PhysRevLett.104.063601,PhysRevA.83.043606}. 

\section{Current measurements}
We now consider the entire system in the presence of  connection between the reservoirs, which we describe with the following Hamiltonian 
\begin{equation}
\hat{H} = \hat{H}_\mathrm{at,L} + \hat{H}_\mathrm{at,R} +  \hat{H}_\mathrm{t} +  \hat{F} \hat{M}, 
\end{equation}
where we introduce a tunneling Hamiltonian $\hat{H}_\mathrm{t}$ \cite{Ingold:1992aa}. We consider the QND regime for the atom-number measurement and replace $\hat{M}$  by $\hat{N}_L/2 $, with $\hat{N}_L$ being the atom number operator in the left reservoir. The Hamiltonians for the left and right reservoirs are identical to $\hat{H}_{at}$ from the previous part, and $\left[ \hat{H}_\mathrm{at,L}, \hat{N}_L \right ] = \left[ \hat{H}_\mathrm{at,R}, \hat{N}_L \right ] = 0$. The Heisenberg equation for $\hat{N}_L$ reduces to the commutator with the tunneling Hamiltonian, and we suppose Kirchoff's law $\dot{\hat{N}}_L = \hat{I}_{at}$, where $\hat{I}_{at}$ is the atomic current operator. 

To describe the back-action of the measurement of $\hat{N}_L$, we introduce the phase operator $\hat{\Phi}$ as the generator of an infinitesimal change of the atom number in the left reservoir \cite{Ingold:1992aa}. By construction, it verifies $\left[ \hat{N}_L, \hat{\Phi} \right ] = i$ and its action on states in the atom-number representation is $-i\frac{\partial}{\partial N_L}$. 
For a closed reservoir without the probe, the phase operator evolves according to $\dot{ \hat{\Phi}} = \frac{\partial \hat{H}_\mathrm{at,L}}{ \partial N_L}$, as a result of $\left[ \hat{H}_\mathrm{at,L}, \hat{N}_L \right ] = 0$. 
Provided that the chemical potential in the right reservoir is fixed, the above equation allows one to identify the fluctuations of $\dot{ \hat{\Phi}}$ as those of the 
chemical-potential difference between the reservoirs. 

The continuous measurement of the atom number yields noise on the phase as a result of Heisenberg's uncertainty principle, which we evaluate using the Heisenberg equation:
\begin{equation}
\dot{\hat{\Phi}} = \frac{1}{i} \left[ \hat{H}_\mathrm{at,L} + \hat{H}_\mathrm{at,R} +  \hat{H}_\mathrm{t} , \hat{\Phi} \right ] + \frac{1}{2}  \hat{F},
\end{equation}
where the first term on the right-hand side is the evolution in the absence of measurement including the dynamical effects of the coupling between the reservoirs and the channel, and the last term represents the random fluctuations due to the continuous observation. Owing to the fact that in the weak-measurement regime, $\hat{F}$ is not correlated to the reservoir dynamics, the power spectrum of the phase fluctuations is given by $\mathcal{S}_{\Phi \Phi}(\omega) = \mathcal{S}_{\Phi \Phi}^0(\omega) + \frac{\mathcal{S}_{FF}(\omega) }{4^2 \omega^2}$, where $\mathcal{S}_{FF}$ is the noise spectrum of $F$, and $ \mathcal{S}_{\Phi \Phi}^0(\omega)$ describes the fluctuations in the absence of measurement~\cite{PhysRevX.6.031002}. For a probe resonant with the cavity in the weak-measurement regime, the dynamical back-action of the measurement vanishes since a small change in the atom number does not alter the intra-cavity photon number. The atomic current noise spectrum is then 
\begin{equation}
\mathcal{S}_{II}(\omega) = \left| Y(\omega) \right|^2 \left(\omega^2 \mathcal{S}_{\Phi \Phi}^0(\omega)  + \frac{\mathcal{S}_{FF}(\omega) }{4} \right),
\end{equation}
where we introduce the frequency-dependent admittance of the channel $Y(\omega)$ and the noise spectral density of $\hat{F}$, $\mathcal{S}_{FF}(\omega)$. This assumes the linear response of the atomic current to small fluctuations around the average bias, but does not  assume the linearity of the current-bias relation itself~\cite{PhysRevX.6.031002}. 

Measurements of the current in the setup of Fig. \ref{fig:twoTerms} will proceed by measurements of the homodyne signal separated in time by $\tau$, yielding the averaged current operator
\begin{equation}
\hat{i}_\tau(t) = \frac{\hat{N}(t+\tau) - \hat{N}(t)}{\tau} = \frac 1 \tau \int_t^{t+\tau}\hat{I}_{at}(u) du,
\end{equation}
where the second equality results from Kirchoff's law. We assume that $\tau$ is much larger than both $1/\kappa$ and the dwell time of atoms in the channel. The total imprecision on the current measurement is then given by the total imprecision originating from detection and the measurement back-action,
\begin{equation}
\mathcal{S}_{ii}^\mathrm{imp}(\omega) =  \mathrm{sinc}^2\left( \frac{\omega \tau}{2} \right) \left[ \frac{\omega ^2 \kappa}{4 g^2}  + \left| Y(\omega) \right|^2 \frac{g^2}{\kappa} \frac{1}{1 + \frac{4\omega^2}{\kappa^2}}\right],
\end{equation}
where consistently with the emergent QND operation we have ignored the equilibrium fluctuations of the atoms within the cavity mode, and we have expressed $\mathcal{S}_{FF}(\omega)$ in terms of the cavity parameters. This expression represents the trade-off between noise and back-action as the measurement strength is varied, similar to the standard quantum limit in cavity optomechanics \cite{Clerk:2010aa,Aspelmeyer:2014aa}.

\begin{figure}[htbp]
\begin{center}
\includegraphics[width=0.35\textwidth]{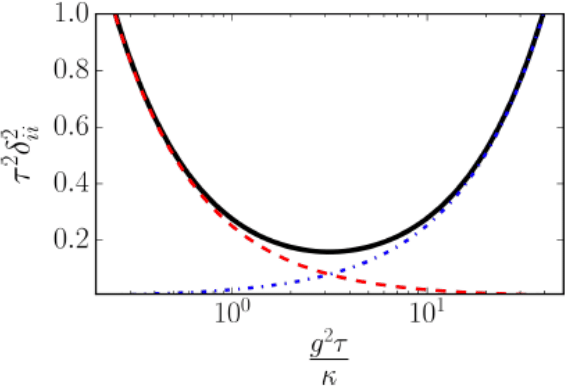}
\caption{(Color online) Total imprecision of current through a fully open quantum point contact over the bandwidth $1/\tau$ as a function of the measurement parameter $g^2 \tau /\kappa$. The dashed red curve represents the contribution of photon shot noise, and the blue dashed-dotted curve represents the contribution from the measurement back-action. }
\label{fig:currentSQL}
\end{center}
\end{figure}

We illustrate this for the case of a fully open quantum point contact at low bias by using the universal conductance quantum as the low-frequency admittance. The total current imprecision $\delta_{ii}^2$ obtained by integration over the bandwidth $1/\tau$ is presented in Fig. \ref{fig:currentSQL}. The lower bound on  current fluctuations is of the order of $1/\tau$, typically two orders of magnitude below the technical noise of the state-of-the-art cold-atom measurements \cite{PhysRevLett.119.030403}. 

\section{Discussion}
The above result is universal in that it does not rely on a Fermi-liquid description of the reservoirs and thus applies to both normal and superfluid phases of interacting fermions. It is a consequence of the existence of the emergent QND measurement of the population of a reservoir, which rests on  Tan's relations. It provides a general framework for the quantum simulation of mesoscopic transport using tunable Fermi gases. This concept differs from other proposals where the current operator couples directly with the cavity field via photon-assisted tunneling, which produces a dissipative current \cite{PhysRevLett.116.060401,PhysRevLett.117.175302,PhysRevA.95.043843}. It also differs from mesosopic electronic devices, in that the two terminals together form a closed system, without electromagnetic environments \cite{PhysRevX.6.031002}, thereby allowing for a simplified and universal analysis.

The most natural experimental platform would consist of cold $^6$Li atoms in the two-terminal configuration accessible in the state-of-the-art experiments~\cite{0953-8984-29-34-343003}, where the light mass of the atom facilitates reaching the QND regime. 
In the presence of a finite $\kappa$ or finite cooperativity, the measurement is not strictly QND. Phenomenologically, we can treat the energy increase in the reservoir as generating a temperature bias across the channel, leading to an extra thermoelectric contribution to the average current \cite{Brantut:2013aa}. 

The QND measurement presented here can be generalized to multi-terminal cases, where a comparable number of cavities or cavity modes monitor several reservoirs simultaneously. Further generalizations could describe  situations, where the cavity is focused on a small region within a single cloud in order to observe the dynamics of the gas \cite{PhysRevLett.120.133601}. In addition, the measurement is, in principle, spin-sensitive such that spin currents as well as particle currents could be monitored along the same principle \cite{Krinner:2016ab}.

\section*{Acknowledments}
We thank C. Galland, T. Donner, and C. Altimiras, T. Giamarchi for discussions and a careful reading of the manuscript, and J. Hofmann and S. Yoshida for discussions on the OPE. JPB acknowledges funding from the ERC project DECCA, EPFL and the Sandoz Family Foundation-Monique de Meuron program for academic promotion.
SU is supported by JSPS KAKENHI Grant No.~JP17K14366.
MU is supported by JSPS KAKENHI Grant
No.~JP26287088, a Grant-in-Aid for Scientic Research on
Innovative Areas ``Topological Materials Science''
(KAKENHI Grant No. JP15H05855), and the Photon Frontier
Network  Program  from  MEXT  of  Japan.

\appendix

\section{Convention}
In the input-output formalism, the interaction between the cavity field $\hat{d}$ and the bath field $\hat{b}$ is modeled as~\cite{gardiner2004}
\beq
H_{\text{int}}=i\int_{-\infty}^{\infty}d\omega\kappa(\omega)
[\hat{d} \hat{b}^{\dagger}(\omega)-\hat{d}^{\dagger}\hat{b}(\omega)],
\eeq
where $\hat{b}$ commutes with all system operators such as those of the cavity and atom fields. When the interaction is of the Markov type, which is of our interest, a frequency dependence of the cavity-bath coupling is neglected and we set
\beq
\kappa(\omega)=\sqrt{\frac{\kappa}{2\pi}}.
\eeq
The input field is defined in terms of the bath field as follows:
\beq
\hat{b}_{\text{in}}(t)=\frac{1}{\sqrt{2\pi}}\int_{-\infty}^{\infty}d\omega e^{-i\omega(t-t_0)}\hat{b}(\omega),
\eeq
where  $t_0$ is an initial time. By using the above convention, the Heisenberg equation of motion for the cavity field is obtained as 
\beq
\dot{\hat{d}}=-i\Omega\hat{M}\hat{d}-\frac{\kappa}{2}\hat{d}-\sqrt{\kappa}\hat{b}_{\text{in}}.
\eeq

In addition, we consider the coherent-state input as described by $\hat{b}_{\text{in}}=-i\sqrt{\Phi}+\hat{\xi}$, where $\hat{\xi}$ describes vacuum-field fluctuations. Therefore, combined with the Markov approximation, the  vacuum fluctuation field satisfies the following relations:
\beq
&&\langle\hat{\xi}(t)\rangle=\langle\hat{\xi}^{\dagger}(t)\rangle=
\langle\hat{\xi}^{\dagger}(t)\hat{\xi}(t')\rangle=\langle\hat{\xi}(t)\hat{\xi}(t')\rangle=0,\nonumber\\
&&\langle\hat{\xi}(t)\hat{\xi}^{\dagger}(t')\rangle=\delta(t-t').
\eeq
In the Fourier space, the above expressions can be rewritten as
\beq
&&\langle\hat{\xi}(\omega)\rangle=\langle\hat{\xi}^{\dagger}(\omega)\rangle
=\langle\hat{\xi}^{\dagger}(\omega)\hat{\xi}(\omega')\rangle
=\langle\hat{\xi}(\omega)\hat{\xi}(\omega')\rangle=0,\nonumber\\
&&\langle\hat{\xi}(\omega)\hat{\xi}^{\dagger}(\omega')\rangle=2\pi\delta(\omega-\omega').
\eeq

\section{Energy absorption rate}
By using the Heisenberg equation of motion, the energy absorption rate is expressed as
\beq
\frac{dE_{\text{at}}(t)}{dt}&&=-i\langle[\hat{H}_{\text{at}}(t),\hat{H}(t)]\rangle
\nonumber\\
&&=-\frac{d}{dt}\langle \hat{M}(t)\hat{F}(t)\rangle
-\frac{\kappa}{2}\langle \hat{M}(t)\hat{F}(t)\rangle\nonumber\\
&&\ \ +2i\Omega\sqrt{\Phi}\langle
\hat{M}(t)(\hat{\xi}(t)-\hat{\xi}^{\dagger}(t))\rangle,
\eeq
where we use
\beq
\frac{d\hat{F}}{dt}=-\frac{\kappa}{2}\hat{F}+2i\Omega\sqrt{\Phi}(\hat{\xi}-\hat{\xi}^{\dagger}).
\label{eq:eq-f}
\eeq
To proceed with the calculation, we next use the condition that $\hat{F}\hat{M}$ is treated as a perturbation. We then obtain
\beq
&&\langle \hat{M}(t)\hat{F}(t)\rangle\approx-i\int_{-\infty}^{t}dt'\langle
           [\hat{M}(t)\hat{F}(t),\hat{M}(t')\hat{F}(t')]\rangle_{0},\nonumber\\\\
&&\langle \hat{M}(t)(\hat{\xi}(t)-\hat{\xi}^{\dagger}(t))\rangle\approx\nonumber\\
&&\ \ -i\int_{-\infty}^{t}dt'\langle
           [\hat{M}(t)(\hat{\xi}(t)-\hat{\xi}^{\dagger}(t)),\hat{M}(t')\hat{F}(t')]\rangle_{0},       
\eeq
where $\langle\cdots\rangle_0$ means the average without $\hat{F}\hat{M}$ in the statistical weight. Since the interaction between atoms and photons is absent without $\hat{F}\hat{M}$, and by assumption there is no initial entanglement between them, the average can be split into photonic and atomic ones, which allows us to obtain a simpler expression. Below, we omit the subscript $0$, as we always calculate averages on the atoms without $\hat{F}\hat{M}$ in the statistical weight.
Equation~\eqref{eq:eq-f} can then be solved as
\beq
\hat{F}(t)=\frac{4i\Omega\sqrt{\Phi}}{\kappa}
[\hat{\tilde{\xi}}(t)-\hat{\tilde{\xi}}^{\dagger}(t)].
\eeq
Here, we introduce
\beq
\hat{\tilde{\xi}}(t)=\int\frac{d\omega}{2\pi}
\frac{e^{-i\omega t}}{1-\frac{2i\omega}{\kappa}}\hat{\xi}(\omega),
\eeq
which accounts for a finite lifetime of photons in the cavity.
Since for $t-t'>0$,
\beq
\langle(\hat{\xi}(t)-\hat{\xi}^{\dagger}(t))(\hat{\tilde{\xi}}(t')-\hat{\tilde{\xi}}^{\dagger}(t'))\rangle
=-\langle\hat{\xi}(t)\hat{\tilde{\xi}}^{\dagger}(t')\rangle\nonumber\\
=-\int\frac{d\omega}{2\pi}\int\frac{d\omega'}{2\pi}
\frac{e^{-i\omega t+i\omega' t'}}{1+\frac{2i\omega'}{\kappa}}
\langle\hat{\xi}(\omega)\hat{\xi}^{\dagger}(\omega')\rangle=0, 
\eeq
and similarly
\beq
\langle(\hat{\tilde{\xi}}(t')-\hat{\tilde{\xi}}^{\dagger}(t'))(\hat{\xi}(t)-\hat{\xi}^{\dagger}(t))\rangle
=0, 
\eeq
we find
\beq
\langle \hat{M}(t)(\hat{\xi}(t)-\hat{\xi}^{\dagger}(t))\rangle=0.
\eeq
The correlation functions between $\hat{F}$'s at different times can be calculated as
\beq
\langle \hat{F}(t)\hat{F}(t')\rangle
&=&\frac{16\Omega^2\Phi}{\kappa^2}\langle \hat{\tilde{\xi}}(t)
\hat{\tilde{\xi}}^{\dagger}(t')
\rangle\nonumber\\
&=&\frac{4\Omega^2\Phi}{\kappa}e^{-\frac{\kappa}{2}(t-t')},
\eeq
and similarly
\beq
\langle \hat{F}(t')\hat{F}(t)\rangle
&=&\frac{4\Omega^2\Phi}{\kappa}e^{-\frac{\kappa}{2}(t-t')},
\eeq
where we again use $t-t'>0$.
Therefore, the energy absorption rate is simplified as
\beq
\frac{dE_{\text{at}}(t)}{dt}
=\frac{4i\Omega^2\Phi}{\kappa}e^{-\frac{\kappa}{2}t}\frac{d}{dt}\left(
\int_{-\infty}^{t}dt'\langle[\hat{M}(t),\hat{M}(t')]\rangle e^{-\frac{\kappa}{2}t'}\right)\nonumber\\.
\eeq
Thus, the problem reduces to the calculation of the average value related to the atomic density.
We note
\beq
&&\langle[\hat{M}(t),\hat{M}(t')]\rangle\nonumber\\
&&=\frac{V}{8}
\int_{-\infty}^{\infty}d\omega e^{-i\omega(t-t')}[S(2k,\omega) 
  -S(2k,-\omega)],
\eeq
where $V$ and $S$ are the volume of the system and the dynamical structure factor, respectively.
To obtain the above result,
we have used the fact that $N$ does not evolve in time, and
\beq
\langle n_{k}(t)n_{k'}(t') \rangle=0,
\eeq
unless $k=-k'$. Equation~(B14) is correct as far as the energy absorption up to $O(\Omega^2)$ is concerned.
In addition, we have also assumed that the system possesses inversion symmetry, which implies
\beq
S(2k,\omega)=S(-2k,\omega).
\eeq
Therefore, the energy absorption rate can be expressed in terms of
the dynamical structure factor as follows:
\beq
\frac{dE_{\text{at}}(t)}{dt}
=\frac{iV\Omega^2\Phi}{4}
     \int_{-\infty}^{\infty}d\omega
     \frac{[S(2k,\omega)-S(2k,-\omega)
     ]}{i\omega+\frac{\kappa}{2}}.
\eeq
     
The above expression contains the integral over the frequency.
To perform the integral, we introduce
the retarded density response function:
\beq
\chi^R(2k,\omega)=\frac{-i}{V}\int_{-\infty}^{\infty} dt
\theta(t)e^{i\omega t}
\langle[n_{2k}(t),n_{-2k}(0)]\rangle.
\eeq
By using the spectral representations for the retarded density response and
dynamical structure factor, 
the following relation is obtained: 
\beq
\text{Im}[\chi^R(2k,\omega)]&&=-\pi[S(2k,\omega)-S(2k,-\omega)]\nonumber\\
&&=
-\text{Im}[\chi^R(-2k,\omega)].
\label{eq:relation-ims}
\eeq
Thus, we obtain
\beq
\frac{dE_{\text{at}}}{dt}
=-\frac{V\Omega^2\Phi}{2\pi}\
     \int_{0}^{\infty}d\omega
     \frac{\omega\text{Im}[\chi^R(2k,\omega)]
     }{\frac{\kappa^2}{4}+\omega^2}.  
\eeq
We next note that the following relation is satisfied:
\beq
\int_0^{\infty}d\omega
\frac{\omega\text{Im}[\chi^R(2k,\omega)]}{\frac{\kappa^2}{4}+\omega^2}=\frac{\pi}{2}
\chi^R(2k,i\kappa/2).
\label{eq:theorem}
\eeq
\begin{figure}[h]
 \begin{center}
  \includegraphics[width=0.35\textwidth]{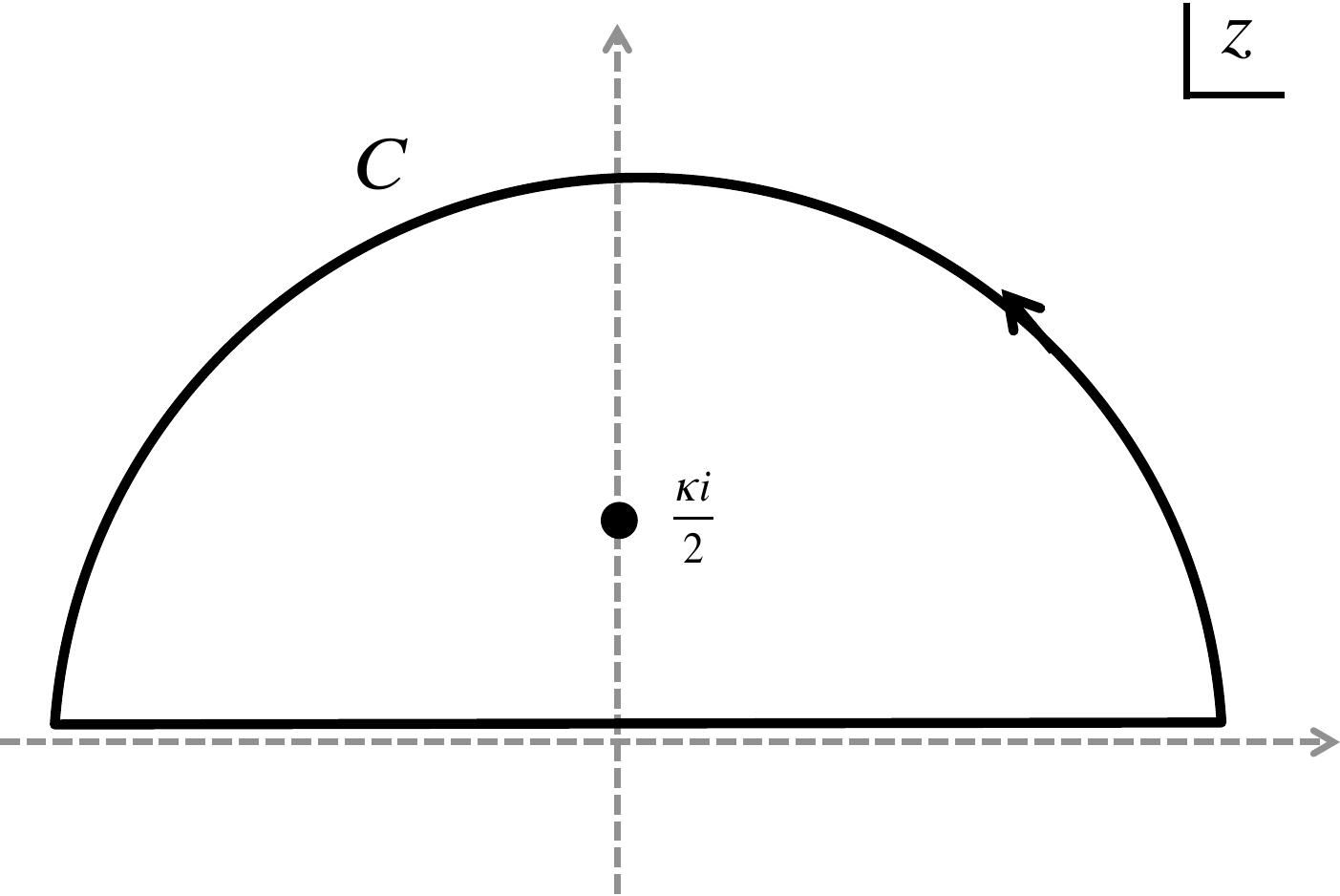}
  \caption{(Color online) Contour used in Eq.~\eqref{eq:contour}. }
  \label{contour}
 \end{center}
\end{figure}
This relation follows from the fact that
the retarded Green's function is analytic for
the upper-half complex plane, $\text{Re}[\chi^R(2k,-\omega)]=\text{Re}[\chi^R(2k,\omega)]$, and
for the contour shown in Fig.~\ref{contour}, we have
\beq
\oint_c \frac{dz}{2\pi i}\frac{z\chi^R(2k,z)}{\frac{\kappa^2}{4}+z^2}
=\frac{1}{2}\chi^R(2k,i\kappa/2).
\label{eq:contour}
\eeq
As a result, the energy absorption rate is given by Eq.~(5).

\section{Density response function at high momentum and frequency}
Here, we discuss the density response function in high-momentum and high-frequency limits. In this case, the operator product expansion (OPE) developed in quantum field theory is available. In the context of two-component Fermi gases, the OPE was originally used to derive Tan's relations~\cite{PhysRevLett.100.205301}. Later on, it was also employed to discuss the asymptotic form of the dynamical structure factor and its sum rules~\cite{PhysRevA.81.063634,PhysRevA.84.043603,PhysRevA.85.013613,PhysRevX.7.011022}.

The OPE states that in the short-range limit a product of two operators $\hat{A}$ and $\hat{B}$ can be expanded in local operators as follows:
\beq
\hat{A}(x)\hat{B}(0)\sim \sum_{i}C_i(x)\hat{O}_i(0),
\label{eq:ope}
\eeq
where $\hat{O}_i$'s are some local operators and $C_i$'s are $c$-number quantities also known as Wilson's coefficients in the OPE. We are interested in the short-range limit of the density response function, where $\hat{A}$ and $\hat{B}$ are given by the density operator. In the Fourier space, this is related to the high-momentum and high-frequency limits of the  density response function:
\beq
\chi^R(q,\omega)=-i\int d^4x e^{iqx}\theta(t)\langle [\hat{n}(x),\hat{n}(0)]\rangle,
\label{eq:density-col}
\eeq
where $qx=\omega t-\mathbf{q}\cdot\mathbf{x}$. To establish Eq.~\eqref{eq:ope} for the density operator, we explicitly calculate \eqref{eq:density-col} for specific few-body states. Such a calculation can be implemented with the Feynman diagrams as discussed in~\cite{PhysRevA.81.063634,PhysRevA.84.043603,PhysRevA.85.013613}. Here, we write down the final expression~\cite{PhysRevA.81.063634,PhysRevA.84.043603,PhysRevA.85.013613}.:
\beq
\chi^R(q,\omega)=c_nn+c_cC
+c_H\mathcal{E}_{\text{at}}n,
\label{eq:density-expansion}
\eeq
where
\beq
n&=&\sum_{\sigma}\langle\hat{\psi}^{\dagger}_{\sigma}\hat{\psi}_{\sigma}\rangle,\\
C&=&m^2U^2\langle \hat{\psi}^{\dagger}_{\uparrow}\hat{\psi}^{\dagger}_{\downarrow}
\hat{\psi}_{\downarrow}\hat{\psi}_{\uparrow}\rangle,\\
\mathcal{E}_{\text{at}}&=&\Big\langle\sum_{\sigma}\hat{\psi}^{\dagger}_{\sigma}\Big(-\frac{\nabla^2}{2m}\Big)\hat{\psi}_{\sigma}+U
 \hat{\psi}^{\dagger}_{\uparrow}\hat{\psi}^{\dagger}_{\downarrow}\hat{\psi}_{\downarrow}\hat{\psi}_{\uparrow}
\Big\rangle/n,
\eeq
with the Fermi field operator $\hat{\psi}$.
Equation~\eqref{eq:density-expansion} shows that  the density correlation function can be expressed in terms of the particle density, Tan's contact density, and energy per atom. Furthermore, their coefficients are given by
\begin{widetext}
\beq
c_n&=&\frac{2\epsilon_q}{((\omega+i\epsilon)^2-\epsilon_{q}^2)},\\
c_c&=&-\frac{1}{m^2}\Big[A(q)\Big\{I_1(q)+\frac{2}{A(0)}
  \frac{1}{\omega-\epsilon_q+i\epsilon}\Big\}^2
  +A(-q)\Big\{I_1(-q)+\frac{2}{A(0)}\frac{1}{-\omega-\epsilon_q-i\epsilon}
  \Big\}^2\nonumber\\
  &&  -\frac{1}{2}\{I_2(q)+I_3(q)+I_2(-q)+I_3(-q)\}\Big]+\frac{2}{m^2A(0)}
\Big(\frac{1}{\omega-\epsilon_q+i\epsilon}+\frac{1}{\omega+\epsilon_q+i\epsilon}\Big)^2\nonumber\\
&&+\frac{4\epsilon_q}{3m^2A(0)}
\Big(\frac{1}{(\omega-\epsilon_q+i\epsilon)^3}-\frac{1}{(\omega+\epsilon_q+i\epsilon
  )^3}\Big),\\
c_H&=&\frac{4\epsilon_q}{3}\Big[\frac{1}{(\omega-\epsilon_q+i\epsilon)^3}
  -\frac{1}{(\omega+\epsilon_q+i\epsilon)^3}\Big],
\eeq
where
\beq
\epsilon_{\mathbf{q}}&=&\frac{\mathbf{q}^2}{2m},\\
A(q)&=&\frac{\frac{4\pi}{m}}{-\frac{1}{a}+\sqrt{-m(\omega-\frac{q^2}{4m})-i\epsilon}},\\
I_1(q)&=&\int \frac{d^3k}{(2\pi)^3}\frac{1}{\epsilon_{\mathbf{k}}}
\frac{1}{\omega-\epsilon_{\mathbf{k}}-\epsilon_{\mathbf{k+q}}+i\epsilon},\\
I_2(q)&=&\int \frac{d^3k}{(2\pi)^3}\frac{1}{\epsilon^2_{\mathbf{k}}}\Big[
  \frac{1}{\omega-\epsilon_{\mathbf{k}}-\epsilon_{\mathbf{k+q}}+i\epsilon}-
  \frac{1}{\omega-\epsilon_{\mathbf{k}}+i\epsilon}\Big],\\
I_3(q)&=&\int \frac{d^3k}{(2\pi)^3}\frac{1}{\epsilon_{\mathbf{k}}\epsilon_{\mathbf{k+q}}}
\frac{1}{\omega-\epsilon_{\mathbf{k}}-\epsilon_{\mathbf{k+q}}+i\epsilon},
\eeq
with an infinitesimal positive parameter $\epsilon$.

To estimate the energy absorption rate, we need $\chi^R$ at pure imaginary frequencies, which can be obtained by the substitution $\omega\to i\omega$ in the above expressions. By using the spectral representation of the correlation function, we can easily show that $\chi^R$ at pure imaginary frequencies becomes real, which ensures that the energy absorption rate also becomes real.

We also note that integrals over momenta for $I_1$, $I_2$, and $I_3$ can be performed analytically. To show the procedure, we consider $I_1$ given by
\beq
I_1(q,i\omega)=-2m^2\int \frac{d^3k}{(2\pi)^3}
\frac{1}{k^2}\frac{1}{k^2+\mathbf{k}\cdot\mathbf{q}+q^2/2-im\omega}.
\eeq
To eliminate the angle dependence of the above expression, we use the Feynman parameter integral,
\beq
\frac{1}{AB}=\int_0^1dx\frac{1}{[Ax+B(1-x)]^2}.
\eeq
Then, we have
\beq
I_1(q,i\omega)
=-2m^2\int_0^1dx\int \frac{d^3k}{(2\pi)^3}
\frac{1}{[k^2-\{x(im\omega-q^2/2)+x^2q^2/4\}]^2}. 
\eeq
By using  the integral formula
\beq
\int \frac{d^3k}{(2\pi)^3}\frac{1}{(k^2-\alpha^2)^2}=\frac{i}{8\pi\alpha},
\ \ \ (\text{Im}[\alpha]>0)
\eeq
we finally obtain
\beq
I_1(q,i\omega)=-\frac{m^2i}{2\pi q}\Big[
2\ln(1+\sqrt{-1+4im\omega/q^2})-\ln(-2+4im\omega/q^2)
  \Big].
\eeq
In a similar manner, we obtain
\beq
I_2(q,i\omega)=-\frac{8m^3i}{\pi q^3}
\frac{\sqrt{-1+4im\omega/q^2}}{(-2+4im\omega/q^2)^2},
\eeq
\beq
I_3(q,i\omega)=\frac{m^3}{2\pi m\omega q}\Big[
  \ln(1-\sqrt{-1+4im\omega/q^2})-3\ln(1+\sqrt{-1+4im\omega/q^2})+\ln(-2+4im\omega/q^2)
  \Big],\nonumber\\
\eeq
and we also find that
$I_j(-q,-i\omega)=I^*_j(q,i\omega)$ ($j=1,2,3$).

Thus, the energy absorption rate per particle is
given by
\beq
\frac{d{\cal E}_{\text{at}}}{dt}=g^2\Bigg[g_n(x)+g_c(x,2ka)\frac{C}{k_F^4}
\left(\frac{k_F}{2k}\right)+g_H(x)\frac{{\cal E}_{\text{at}}}{{\cal E}_{0\text{at}}}
\left(\frac{k_F}{2k}\right)^2
\Bigg],
\eeq
where
\beq
g_n(x)&=&\frac{1}{8}\frac{x}{x^2+1},
\eeq
\beq
g_c(x,2ka)&=&-\frac{3\pi}{16}
\frac{x}{-\frac{1}{2ka}+\sqrt{\frac{1}{4}-\frac{ix}{2}}}
     [2\ln(1+\sqrt{-1+2ix})-\ln(-2+2ix)]^2\nonumber\\
&&-\frac{3\pi}{16}
\frac{x}{-\frac{x}{2ka}+\sqrt{\frac{1}{4}+\frac{ix}{2}}}
     [2\ln(1+\sqrt{-1-2ix})-\ln(-2-2ix)]^2 \nonumber\\
    && +\frac{3\pi}{4}\frac{1}{1-2ka\sqrt{\frac{1}{4}-\frac{ix}{2}}}\frac{ix}{1-ix}
     [2\ln(1+\sqrt{-1+2ix})-\ln(-2+2ix)]\nonumber\\
    && -\frac{3\pi}{4}\frac{1}{1-2ka\sqrt{\frac{1}{4}+\frac{ix}{2}}}\frac{ix}{1+ix}
     [2\ln(1+\sqrt{-1-2ix})-\ln(-2-2ix)]\nonumber\\
    && -\frac{3\pi}{4}\frac{1}{2ka-(2ka)^2\sqrt{\frac{1}{4}-\frac{ix}{2}}} \frac{x}{(1-ix)^2} 
     -\frac{3\pi}{4}\frac{1}{2ka-(2ka)^2\sqrt{\frac{1}{4}+\frac{ix}{2}}} \frac{x}{(1+ix)^2}
     \nonumber\\
    && -\frac{3\pi}{32}[\ln(1-\sqrt{-1+2ix})-3\ln(1+\sqrt{-1+2ix})+\ln(-2+2ix)]
     \nonumber\\
    && -\frac{3\pi}{32}[\ln(1-\sqrt{-1-2ix})-3\ln(1+\sqrt{-1-2ix})+\ln(-2-2ix)]
    \nonumber\\
    && +\frac{3i\pi }{16}\frac{x\sqrt{-1+2ix}}{(-1+ix)^2}
     -\frac{3i\pi }{16}\frac{x\sqrt{-1-2ix}}{(-1-ix)^2}
     -\frac{3\pi}{4ka}\frac{x^3}{(x^2+1)^2}
     -\frac{\pi}{4ka}\frac{x(1-3x^2)}{(x^2+1)^3},
\eeq
\beq
g_H(x)&=&\frac{1}{5}\frac{x(1-3x^2)}{(x^2+1)^3}.
\eeq

\section{noise of the homodyne photocurrent}
We derive the noise expression of the homodyne photocurrent. To this end, we consider 
\beq
\langle \hat{I}_h(t)\hat{I}_h(0)\rangle
&=&\Big\langle\Big[\hat{\xi}(t)+\hat{\xi}^{\dagger}(t)+\frac{8\Omega\sqrt{\Phi}}{\kappa}\hat{\tilde{M}}(t)
  -2(\hat{\tilde{\xi}}(t)+\hat{\tilde{\xi}}^{\dagger}(t))\Big]\nonumber\\
&&\times\Big[\hat{\xi}(0)+\hat{\xi}^{\dagger}(0)+\frac{8\Omega\sqrt{\Phi}}{\kappa}\hat{\tilde{M}}(0)
  -2(\hat{\tilde{\xi}}(0)+\hat{\tilde{\xi}}^{\dagger}(0))\Big]\Big\rangle.
\eeq
It is then straightforward to calculate the following correlation functions between the noise fields and between atoms:
\beq
\langle(\hat{\xi}(t)+\hat{\xi}^{\dagger}(t))(\hat{\xi}(0)+\hat{\xi}^{\dagger}(0))\rangle=\delta(t),
\eeq
\beq
4\langle(\hat{\tilde{\xi}}(t)+\hat{\tilde{\xi}}^{\dagger}(t))
(\hat{\tilde{\xi}}(0)+\hat{\tilde{\xi}}^{\dagger}(0))\rangle
=\kappa e^{-\frac{\kappa}{2}|t|},
\eeq
\beq
-2\langle(\hat{\tilde{\xi}}(t)+\hat{\tilde{\xi}}^{\dagger}(t))(\hat{\xi}(0)+\hat{\xi}^{\dagger}(0))\rangle
-\kappa e^{\frac{\kappa}{2}t}\theta(t),
\eeq
\beq
-2\langle(\hat{\xi}(t)+\hat{\xi}^{\dagger}(t))(\hat{\tilde{\xi}}(0)+\hat{\tilde{\xi}}^{\dagger}(0))\rangle
=-\kappa e^{\frac{\kappa}{2}t}\theta(-t),
\eeq
\beq
\frac{64\Omega^2\Phi}{\kappa^2}\langle\hat{\tilde{M}}(t)\hat{\tilde{M}}(0)\rangle
=\frac{16\Omega^2\Phi N^2}{\kappa^2}+\frac{8\Omega^2\Phi V}{\kappa^2}
\int d\omega
\frac{e^{-i\omega t}}{1+\frac{4\omega^2}{\kappa^2}}S(2k,\omega).
\eeq
Thus, the correlation function is simplified as
\beq
\langle \hat{I}_h(t)\hat{I}_h(0)\rangle=\delta(t)
+\frac{16\Omega^2\Phi N^2}{\kappa^2}+\frac{8\Omega^2\Phi V}{\kappa^2}
\int_{-\infty}^{\infty}d\omega\frac{e^{-i\omega t}}{1+\frac{4\omega^2}{\kappa^2}}S(2k,\omega)\nonumber\\
+\frac{8\Omega\sqrt{\Phi}}{\kappa}\langle (\hat{\xi}(t)+\hat{\xi}^{\dagger}(t))
\hat{\tilde{M}}(0)\rangle
+\frac{8\Omega\sqrt{\Phi}}{\kappa}\langle\hat{\tilde{M}}(t)(\hat{\xi}(0)+\hat{\xi}^{\dagger}(0))\rangle\nonumber\\
-\frac{16\Omega\sqrt{\Phi}}{\kappa}\langle (\hat{\tilde{\xi}}(t)+\hat{\tilde{\xi}}^{\dagger}(t))\hat{\tilde{M}}(0)\rangle
-\frac{16\Omega\sqrt{\Phi}}{\kappa}\langle\hat{\tilde{M}}(t)(\hat{\tilde{\xi}}(0)+\hat{\tilde{\xi}}^{\dagger}(0))\rangle.
\eeq
We use linear response theory to calculate the remaining correlations, which are summarized as follows:
\beq
\frac{8\Omega\sqrt{\Phi}}{\kappa}\langle (\hat{\xi}(t)+\hat{\xi}^{\dagger}(t))\hat{\tilde{M}}(0)\rangle
=\frac{4\Omega^2\Phi V}{\kappa^2}\theta(-t)
\int_{-\infty}^{\infty} d\omega
\frac{[e^{\frac{\kappa}{2}t}-e^{i\omega t}][S(2k,\omega)-S(2k,-\omega)]}{(1-\frac{2i\omega}{\kappa})^2},
\eeq
\beq
\frac{8\Omega\sqrt{\Phi}}{\kappa}\langle\hat{\tilde{M}}(t)
(\hat{\xi}(0)+\hat{\xi}^{\dagger}(0))\rangle
\approx-\frac{4\Omega^2\Phi V}{\kappa^2}\theta(t)
\int_{-\infty}^{\infty} d\omega 
\frac{[e^{-\frac{\kappa}{2}t}-e^{-i\omega t}][S(2k,\omega)-S(2k,-\omega)]}{(1-\frac{2i\omega}{\kappa})^2},
\eeq
\beq
-\frac{16\Omega\sqrt{\Phi}}{\kappa}\langle
(\hat{\tilde{\xi}}(t)+\hat{\tilde{\xi}}^{\dagger}(t))\hat{\tilde{M}}(0)\rangle
\approx\frac{4\Omega^2\Phi V}{\kappa^2}\theta(-t)\int_{-\infty}^{\infty} d\omega
\Big[\frac{e^{i\omega t}[S(2k,\omega)-S(2k,-\omega)]}{1+\frac{4\omega^2}{\kappa^2}}\nonumber\\
-\frac{[e^{\frac{\kappa}{2}t}-e^{i\omega t}][S(2k,\omega)-S(2k,-\omega)]}{(1-\frac{2i\omega}{\kappa})^2}  
\Big],
\eeq
\beq
-\frac{16\Omega\sqrt{\Phi}}{\kappa}\langle\hat{\tilde{M}}(t)
(\hat{\tilde{\xi}}(0)+\hat{\tilde{\xi}}^{\dagger}(0))\rangle
\approx-\frac{4\Omega^2\Phi V}{\kappa^2}\theta(t)\int_{-\infty}^{\infty} d\omega
\Big[\frac{e^{-i\omega t}[S(2k,\omega)-S(2k,-\omega)]}{1+\frac{4\omega^2}{\kappa^2}}\nonumber\\
  - \frac{[e^{-\frac{\kappa}{2}t}-e^{-i\omega t}][S(2k,\omega)-S(2k,-\omega)]
  }{(1-\frac{2i\omega}{\kappa})^2}\Big].  
\eeq
Thus, the noise spectral density is obtained as
\beq
\langle I_{h}(t)I_h(0)\rangle=
\delta(t)+\frac{4g^2N^2}{\kappa}+\frac{g^2V}{\kappa}\int_{-\infty}^{\infty}d\omega
\frac{e^{-i\omega t}}{1+\frac{4\omega^2}{\kappa^2}}[S(2k,\omega)+S(2k,-\omega)].
\eeq
It follows from this result that the noise spectral density referred back to the atom number is given in Eq.~(6).

We finally discuss how the dynamical structure factor behaves under $2k\gg k_F$. Since the dynamical structure factor has the same dimension as the density response function, it is useful to multiply it by $(2k)^2/(2mn)$. Then, by means of the OPE, the dimensionless dynamical structure factor $\bar{S}$ can be obtained as~\cite{PhysRevA.81.063634,PhysRevA.84.043603,PhysRevA.85.013613}
\beq
\bar{S}(2k,\omega)\equiv\frac{(2k)^2}{2mn}\times S(2k,\omega)=\Big(\frac{C}{k_F^4}\Big)\Big(\frac{k_F}{2k}\Big)
\Bigg[\theta(y-\frac{1}{2})\Big\{\frac{3\sqrt{2}}{2}
  \frac{\sqrt{y-\frac{1}{2}}}{(y-1)^2}
  +\frac{3}{2y}
  \log\Big(\frac{y+\sqrt{2}\sqrt{y-\frac{1}{2}}}{|y-1|}\Big)
  \Big\}\nonumber\\
  -\frac{3\sqrt{2}}{2}\text{Im}\Big[
    \frac{1}{-\frac{\sqrt{2}}{2ka}+\sqrt{-y+\frac{1}{2}-i\eta}}
    \Big\{-i \log\Big(\frac{y+\sqrt{2}\sqrt{y-\frac{1}{2}}}{|y-1|}\Big)
    -\pi\theta(1-y)-\frac{2}{2ka(y-1)}
    \Big\}^2
    \Big]
  \Bigg],
\eeq
where $y=\frac{m\omega}{2k^2}$. We note that the dynamical structure factor obtained above diverges at the single particle peak $y=1$, which is known to be an artifact of the OPE~\cite{PhysRevA.81.063634}. More recently, it has been pointed out in Ref.~\cite{PhysRevX.7.011022} that such an artifact can be resolved by considering an impulse approximation that is correct near the single-particle peak at $y=1$. Under the impulse approximation, the dynamical structure factor is expressed as
\beq
S_{IA}(2k,\omega)=\int\frac{d^3k}{(2\pi)^3}n(\mathbf{k})\delta(\omega+\epsilon_{\mathbf{k}}-\epsilon_{\mathbf{k+2k}}).
\eeq
\end{widetext}
As shown in Fig.~5, global behaviors of the dimensionless dynamical structure factor can be obtained by connecting the OPE and impulse approximation expressions in a smooth manner~\cite{PhysRevX.7.011022}. 
\begin{figure}[h]
 \begin{center}
  \includegraphics[width=0.35\textwidth]{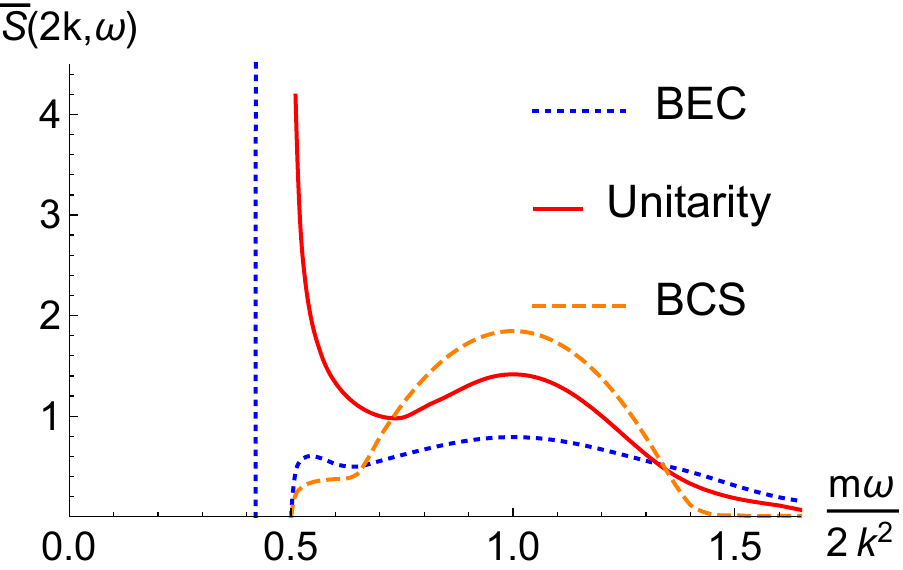}
  \caption{(Color online) Typical behaviors of the dynamical structure factor. The dashed orange, red, and dotted blue curves
  represent the behaviors in the BCS ($1/(k_Fa)=-1$), unitarity ($1/(k_Fa)=0$),
  and BEC regime ($1/(k_Fa)=1$), respectively. }
 \end{center}
\end{figure}

\section{Comparison with a ring cavity}

To study the role of the particular mode function in the noise measurement and heating rate, we now reproduce the reasoning leading to equation~(A4) for the case of a ring cavity, which comprises two counterpropagating, degenerate modes, which we label $\hat{d}_+$ and $\hat{d}_-$. The dispersive coupling between the atoms and the field is written as

\begin{equation}
\hat{H}_{\mathrm{At-Field}} = \Omega \int d\mathbf{r} \left| e^{ikz} \hat{d}_+ + e^{-ikz} \hat{d}_- \right|^2 \hat{n}(\mathbf{r},t).
\end{equation}

Expanding the field strength, and introducing the Fourier components of the density, we obtain
\begin{equation}
\hat{H}_{\mathrm{At-Field}}  =  \Omega \left( \hat{d}_+^\dagger \hat{d}_+ +  \hat{d}_-^\dagger \hat{d}_- \right) \hat{N} + \Omega \left(\hat{d}_-^\dagger  \hat{d}_+ \hat{n}_{-2k} + \hat{d}_+^\dagger \hat{d}_- \hat{n}_{2k} \right).
\end{equation}

As in the case of the Fabry-Perot cavity,
the Heisenberg equations of motion for the photon fields under resonant driving can be obtained as
\beq
\dot{\hat{d}}_+(t) &=&  -i\Omega {N} \hat{d}_+ -\frac \kappa 2 \hat{d}_+ - \sqrt{\kappa} \hat{b}_{+,\mathrm{in}} - i\Omega \hat{d}_- \hat{n}_{2k}, \\
\dot{\hat{d}}_-(t) &=&  -i\Omega {N} \hat{d}_- -\frac \kappa 2 \hat{d}_- - \sqrt{\kappa} \hat{b}_{-,\mathrm{in}} - i\Omega \hat{d}_+ \hat{n}_{-2k},\nonumber\\
\eeq
where we have introduced the input modes in the '$+$' and '$-$' directions as $\hat{b}_{+,\mathrm{in}}$ and $\hat{b}_{-,\mathrm{in}}$, respectively.
In order to apply the measurement protocol, we drive the cavity resonantly with a coherent state $\bar{b}_{+,\mathrm{in}}$ in the cavity in the '$+$' mode, while the '$-$' mode remains in the vacuum state. We introduce the vacuum noise operators and the coherent states in each modes by
\beq
\hat{d}_+(t)                   &=&  \bar{d}_+ + \hat{\eta}_+(t),\\
\hat{d}_-(t)                    &=&  \bar{d}_- + \hat{\eta}_-(t), \\
\hat{b}_{+,\mathrm{in}} &=&  \bar{b}_{+,\mathrm{in}} + \hat{\xi}_+,\\
\hat{b}_{+,\mathrm{in}} &=&  \hat{\xi}_-.
\eeq

As in the Fabry-Perot cavity case, we treat $\Omega \hat{n}_i$ and the vacuum noise amplitudes as small quantities. With $\bar{b}_+ = -i \sqrt{\Phi}$ and to zeroth order, we obtain $\bar{d}_+ =  2i\sqrt{\frac \Phi \kappa}$ and $\bar{d}_- = 0$.
By using this zeroth-order result, we obtain the following first-order equations:
\beq
\dot{\hat{\eta}}_+(t) &=&  -\frac \kappa 2 \hat{\eta}_+ - \sqrt{\kappa} \hat{\xi}_+
+g\hat{N},\\
\dot{\hat{\eta}}_-(t) &=&   -\frac \kappa 2 \hat{\eta}_-  - \sqrt{\kappa} \hat{\xi}_-
+g\hat{n}_{-2k}.
\eeq
We observe that the noise on the '$+$' mode up to this order is not coupled to $n_{2k}$ due to the
absence of coherent driving in $\hat{b}_{-,\text{in}}$. Therefore a measurement of the outgoing field in the forward direction does not carry any extra noise apart from shot noise, in contrast to the case of a standing-wave mode treated in the Fabry-Perot cavity case.

To understand heating due to the vacuum fluctuations entering the cavity, we rewrite the coupling Hamiltonian in terms of the fluctuating fields.
As in the case of the Fabry-Perot cavity, we focus on the part of the Hamiltonian that couples the light with the atomic density fluctuations, which is given by 
\begin{equation}
\hat{H}_{\mathrm{At-Field}}  
 =ig\left( \hat{\eta}_+^\dagger - \hat{\eta}_+  \right) \hat{N}
 +i g  \left( \hat{\eta}_-^\dagger \hat{n}_{-2k} - \hat{\eta}_- \hat{n}_{2k} \right).
\end{equation}
The energy absorption rate in the presence of $\hat{H}_{\mathrm{At-Field}} $
can be calculated in a manner similar to the Fabry-Perot cavity, within second-order perturbation theory. 
We find that the energy absorption rate per particle in the ring cavity case is obtained as
\beq
\frac{d{\cal E}_{\text{at}}}{dt}=-\frac{g^2\kappa }{2n}\chi^R(2k,i\kappa/2).
\eeq
We note that the above expression is similar to one for the Fabry-Perot cavity except for
the factor of $1/8$.
The difference in the factor can easily be understood by recalling that
there are the factor $1/4$ in front of $\hat{n}_{\pm2k}$ and the $S(-2k,\omega)$
contribution from the Fabry-Perot cavity.

\section{Spontaneous emission}

We consider the effects of spontaneous emission in the regime of a dispersive coupling of the cavity to the atoms, modeled as two-level systems, where the detuning $\Delta$ of the cavity with respect to the atomic resonance is very large compared with the natural decay of the excited state $\gamma$. The presence of spontaneous emission is effectively accounted for by a modified Schr\"odinger equation for the wave function of the atom projected onto the excited state $\psi_e$:
\begin{equation}
\dot{\psi_e} = i \left[ \frac{\nabla^2}{2m} + \Delta\right] \psi_e - i g_0 \hat{d} \psi_g - \gamma \psi_e,
\end{equation}
where $\psi_g$ is the wave function of the atom in the ground state, and $g_0$ is twice the single-photon Rabi frequency. Here we only consider spontaneous emission at the single-atom level. 
For large detuning, the rate of spontaneous emission in free space is then
\begin{equation}
\Gamma = \frac{4 \Phi}{\kappa} \frac{g_0^2 \gamma}{\Delta^2},
\end{equation}
and the dispersive shift is 
\begin{equation}
\Omega = \frac{g_0^2 }{\Delta}.
\end{equation}

\begin{figure}[htb]
 \begin{center}
  \includegraphics[width=0.35\textwidth]{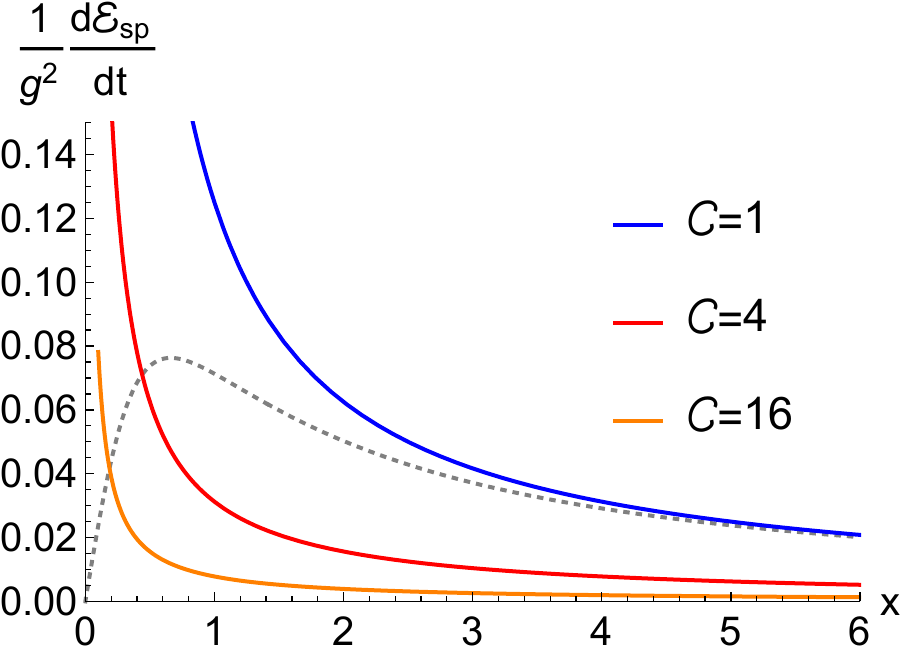}
  \caption{(Color online) Heating rate due to spontaneous emission  as a function of $x$ for different values of the cooperativity. For comparison the heating rate due to measurement back-action for the unitary Fermi gas is shown as a dashed curve.}
  \label{fig:heatingSupp}
 \end{center}
\end{figure}

Considering the fact that the average energy of an atom undergoing  spontaneous emission is entirely dissipated in the cloud, the average increase of the energy per particle due to spontaneous emission is then 
\begin{equation}
\frac{d \mathcal{E}_\mathrm{sp}}{dt} = \Gamma \epsilon_r,
\end{equation}
where $\epsilon_r = k^2/2m$ is the recoil energy \cite{Gerbier:2010aa}. This is actually an upper bound since in many cases the atom will not dissipate its energy in the cloud but will rather be lost, yielding a heating of the order of the Fermi energy rather than the recoil energy. With the notations used in $\mathcal{E}_{\mathrm{at}}$, we obtain 
\begin{equation}
\frac{1}{ g^2} \frac{d \mathcal{E}_\mathrm{sp}}{dt}  = \frac{\gamma}{g_0^2}\epsilon_r.
\end{equation}
Introducing the cooperativity of the cavity $\mathcal{C} = g_0^2/\kappa\gamma$, we obtain
\begin{equation}
\frac{1}{ g^2} \frac{d \mathcal{E}_\mathrm{sp}}{dt}  = \frac{1}{ 8x \mathcal{C}}.
\end{equation}

Figure \ref{fig:heatingSupp} presents the heating rate due to spontaneous emission as a function of $x$ for several  cooperativities. Note that due to the normalisation used, this is also done at a fixed photon number in the cavity. We observe that even for moderate cooperativities, the measurement back-action dominates over spontaneous emission over a significant range of the bandwidth.


\begin{thebibliography}{48}%
\makeatletter
\providecommand \@ifxundefined [1]{%
 \@ifx{#1\undefined}
}%
\providecommand \@ifnum [1]{%
 \ifnum #1\expandafter \@firstoftwo
 \else \expandafter \@secondoftwo
 \fi
}%
\providecommand \@ifx [1]{%
 \ifx #1\expandafter \@firstoftwo
 \else \expandafter \@secondoftwo
 \fi
}%
\providecommand \natexlab [1]{#1}%
\providecommand \enquote  [1]{``#1''}%
\providecommand \bibnamefont  [1]{#1}%
\providecommand \bibfnamefont [1]{#1}%
\providecommand \citenamefont [1]{#1}%
\providecommand \href@noop [0]{\@secondoftwo}%
\providecommand \href [0]{\begingroup \@sanitize@url \@href}%
\providecommand \@href[1]{\@@startlink{#1}\@@href}%
\providecommand \@@href[1]{\endgroup#1\@@endlink}%
\providecommand \@sanitize@url [0]{\catcode `\\12\catcode `\$12\catcode
  `\&12\catcode `\#12\catcode `\^12\catcode `\_12\catcode `\%12\relax}%
\providecommand \@@startlink[1]{}%
\providecommand \@@endlink[0]{}%
\providecommand \url  [0]{\begingroup\@sanitize@url \@url }%
\providecommand \@url [1]{\endgroup\@href {#1}{\urlprefix }}%
\providecommand \urlprefix  [0]{URL }%
\providecommand \Eprint [0]{\href }%
\providecommand \doibase [0]{http://dx.doi.org/}%
\providecommand \selectlanguage [0]{\@gobble}%
\providecommand \bibinfo  [0]{\@secondoftwo}%
\providecommand \bibfield  [0]{\@secondoftwo}%
\providecommand \translation [1]{[#1]}%
\providecommand \BibitemOpen [0]{}%
\providecommand \bibitemStop [0]{}%
\providecommand \bibitemNoStop [0]{.\EOS\space}%
\providecommand \EOS [0]{\spacefactor3000\relax}%
\providecommand \BibitemShut  [1]{\csname bibitem#1\endcsname}%
\let\auto@bib@innerbib\@empty
\bibitem [{\citenamefont {Cirac}\ and\ \citenamefont
  {Zoller}(2012)}]{Cirac:2012aa}%
  \BibitemOpen
  \bibfield  {author} {\bibinfo {author} {\bibfnamefont {J.~I.}\ \bibnamefont
  {Cirac}}\ and\ \bibinfo {author} {\bibfnamefont {P.}~\bibnamefont {Zoller}},\
  }\href@noop {} {\bibfield  {journal} {\bibinfo  {journal} {Nature Physics}\
  }\textbf {\bibinfo {volume} {8}},\ \bibinfo {pages} {264} (\bibinfo {year}
  {2012})}\BibitemShut {NoStop}%
\bibitem [{\citenamefont {Chien}\ \emph {et~al.}(2015)\citenamefont {Chien},
  \citenamefont {Peotta},\ and\ \citenamefont {Di~Ventra}}]{Chien:2015ab}%
  \BibitemOpen
  \bibfield  {author} {\bibinfo {author} {\bibfnamefont {C.-C.}\ \bibnamefont
  {Chien}}, \bibinfo {author} {\bibfnamefont {S.}~\bibnamefont {Peotta}}, \
  and\ \bibinfo {author} {\bibfnamefont {M.}~\bibnamefont {Di~Ventra}},\ }\href
  {http://dx.doi.org/10.1038/nphys3531} {\bibfield  {journal} {\bibinfo
  {journal} {Nat Phys}\ }\textbf {\bibinfo {volume} {11}},\ \bibinfo {pages}
  {998} (\bibinfo {year} {2015})}\BibitemShut {NoStop}%
\bibitem [{\citenamefont {Krinner}\ \emph {et~al.}(2017)\citenamefont
  {Krinner}, \citenamefont {Esslinger},\ and\ \citenamefont
  {Brantut}}]{0953-8984-29-34-343003}%
  \BibitemOpen
  \bibfield  {author} {\bibinfo {author} {\bibfnamefont {S.}~\bibnamefont
  {Krinner}}, \bibinfo {author} {\bibfnamefont {T.}~\bibnamefont {Esslinger}},
  \ and\ \bibinfo {author} {\bibfnamefont {J.-P.}\ \bibnamefont {Brantut}},\
  }\href {http://stacks.iop.org/0953-8984/29/i=34/a=343003} {\bibfield
  {journal} {\bibinfo  {journal} {Journal of Physics: Condensed Matter}\
  }\textbf {\bibinfo {volume} {29}},\ \bibinfo {pages} {343003} (\bibinfo
  {year} {2017})}\BibitemShut {NoStop}%
\bibitem [{\citenamefont {Blanter}\ and\ \citenamefont
  {Buttiker}(2000)}]{BLANTER20001}%
  \BibitemOpen
  \bibfield  {author} {\bibinfo {author} {\bibfnamefont {Y.}~\bibnamefont
  {Blanter}}\ and\ \bibinfo {author} {\bibfnamefont {M.}~\bibnamefont
  {Buttiker}},\ }\href {\doibase https://doi.org/10.1016/S0370-1573(99)00123-4}
  {\bibfield  {journal} {\bibinfo  {journal} {Physics Reports}\ }\textbf
  {\bibinfo {volume} {336}},\ \bibinfo {pages} {1 } (\bibinfo {year}
  {2000})}\BibitemShut {NoStop}%
\bibitem [{\citenamefont {Lye}\ \emph {et~al.}(2003)\citenamefont {Lye},
  \citenamefont {Hope},\ and\ \citenamefont {Close}}]{Lye:2003aa}%
  \BibitemOpen
  \bibfield  {author} {\bibinfo {author} {\bibfnamefont {J.~E.}\ \bibnamefont
  {Lye}}, \bibinfo {author} {\bibfnamefont {J.~J.}\ \bibnamefont {Hope}}, \
  and\ \bibinfo {author} {\bibfnamefont {J.~D.}\ \bibnamefont {Close}},\ }\href
  {\doibase 10.1103/PhysRevA.67.043609} {\bibfield  {journal} {\bibinfo
  {journal} {Phys. Rev. A}\ }\textbf {\bibinfo {volume} {67}},\ \bibinfo
  {pages} {043609} (\bibinfo {year} {2003})}\BibitemShut {NoStop}%
\bibitem [{\citenamefont {Tanji-Suzuki}\ \emph {et~al.}(2011)\citenamefont
  {Tanji-Suzuki}, \citenamefont {Leroux}, \citenamefont {Schleier-Smith},
  \citenamefont {Cetina}, \citenamefont {Grier}, \citenamefont {Simon},\ and\
  \citenamefont {Vuletic}}]{2011AAMOP..60..201T}%
  \BibitemOpen
  \bibfield  {author} {\bibinfo {author} {\bibfnamefont {H.}~\bibnamefont
  {Tanji-Suzuki}}, \bibinfo {author} {\bibfnamefont {I.~D.}\ \bibnamefont
  {Leroux}}, \bibinfo {author} {\bibfnamefont {M.~H.}\ \bibnamefont
  {Schleier-Smith}}, \bibinfo {author} {\bibfnamefont {M.}~\bibnamefont
  {Cetina}}, \bibinfo {author} {\bibfnamefont {A.~T.}\ \bibnamefont {Grier}},
  \bibinfo {author} {\bibfnamefont {J.}~\bibnamefont {Simon}}, \ and\ \bibinfo
  {author} {\bibfnamefont {V.}~\bibnamefont {Vuletic}},\ }\href@noop {}
  {\bibfield  {journal} {\bibinfo  {journal} {Advances in Atomic Molecular and
  Optical Physics}\ }\textbf {\bibinfo {volume} {60}},\ \bibinfo {pages} {201}
  (\bibinfo {year} {2011})}\BibitemShut {NoStop}%
 \bibitem [{\citenamefont {Tan}(2008)}]{Tan:2008ac}%
  \BibitemOpen
  \bibfield  {author} {\bibinfo {author} {\bibfnamefont {S.}~\bibnamefont
  {Tan}},\ }\href@noop {} {\bibfield  {journal} {\bibinfo  {journal} {Annals of
  Physics}\ }\textbf {\bibinfo {volume} {323}},\ \bibinfo {pages} {2971}
  (\bibinfo {year} {2008})}\BibitemShut {NoStop}%
\bibitem [{\citenamefont {Braaten}(2012)}]{braaten2012universal}%
  \BibitemOpen
  \bibfield  {author} {\bibinfo {author} {\bibfnamefont {E.}~\bibnamefont
  {Braaten}},\ }in\ \href@noop {} {\emph {\bibinfo {booktitle} {The BCS-BEC
  Crossover and the Unitary Fermi Gas}}}\ (\bibinfo  {publisher} {Springer},\
  \bibinfo {year} {2012})\ pp.\ \bibinfo {pages} {193--231}\BibitemShut
  {NoStop}%
\bibitem [{\citenamefont {{Eckert}}\ \emph {et~al.}(2008)\citenamefont
  {{Eckert}}, \citenamefont {{Romero-Isart}}, \citenamefont {{Rodriguez}},
  \citenamefont {{Lewenstein}}, \citenamefont {{Polzik}},\ and\ \citenamefont
  {{Sanpera}}}]{Eckert:2008aa}%
  \BibitemOpen
  \bibfield  {author} {\bibinfo {author} {\bibfnamefont {K.}~\bibnamefont
  {{Eckert}}}, \bibinfo {author} {\bibfnamefont {O.}~\bibnamefont
  {{Romero-Isart}}}, \bibinfo {author} {\bibfnamefont {M.}~\bibnamefont
  {{Rodriguez}}}, \bibinfo {author} {\bibfnamefont {M.}~\bibnamefont
  {{Lewenstein}}}, \bibinfo {author} {\bibfnamefont {E.~S.}\ \bibnamefont
  {{Polzik}}}, \ and\ \bibinfo {author} {\bibfnamefont {A.}~\bibnamefont
  {{Sanpera}}},\ }\href {\doibase 10.1038/nphys776} {\bibfield  {journal}
  {\bibinfo  {journal} {Nature Physics}\ }\textbf {\bibinfo {volume} {4}},\
  \bibinfo {pages} {50} (\bibinfo {year} {2008})},\ \Eprint
  {http://arxiv.org/abs/0709.0527} {arXiv:0709.0527} \BibitemShut {NoStop}%
\bibitem [{\citenamefont {Hauke}\ \emph {et~al.}(2013)\citenamefont {Hauke},
  \citenamefont {Sewell}, \citenamefont {Mitchell},\ and\ \citenamefont
  {Lewenstein}}]{PhysRevA.87.021601}%
  \BibitemOpen
  \bibfield  {author} {\bibinfo {author} {\bibfnamefont {P.}~\bibnamefont
  {Hauke}}, \bibinfo {author} {\bibfnamefont {R.~J.}\ \bibnamefont {Sewell}},
  \bibinfo {author} {\bibfnamefont {M.~W.}\ \bibnamefont {Mitchell}}, \ and\
  \bibinfo {author} {\bibfnamefont {M.}~\bibnamefont {Lewenstein}},\ }\href
  {\doibase 10.1103/PhysRevA.87.021601} {\bibfield  {journal} {\bibinfo
  {journal} {Phys. Rev. A}\ }\textbf {\bibinfo {volume} {87}},\ \bibinfo
  {pages} {021601} (\bibinfo {year} {2013})}\BibitemShut {NoStop}%
\bibitem [{\citenamefont {Lee}\ and\ \citenamefont
  {Ruostekoski}(2014)}]{PhysRevA.90.023628}%
  \BibitemOpen
  \bibfield  {author} {\bibinfo {author} {\bibfnamefont {M.~D.}\ \bibnamefont
  {Lee}}\ and\ \bibinfo {author} {\bibfnamefont {J.}~\bibnamefont
  {Ruostekoski}},\ }\href {\doibase 10.1103/PhysRevA.90.023628} {\bibfield
  {journal} {\bibinfo  {journal} {Phys. Rev. A}\ }\textbf {\bibinfo {volume}
  {90}},\ \bibinfo {pages} {023628} (\bibinfo {year} {2014})}\BibitemShut
  {NoStop}%
\bibitem [{\citenamefont {Mekhov}\ and\ \citenamefont
  {Ritsch}(2009)}]{PhysRevLett.102.020403}%
  \BibitemOpen
  \bibfield  {author} {\bibinfo {author} {\bibfnamefont {I.~B.}\ \bibnamefont
  {Mekhov}}\ and\ \bibinfo {author} {\bibfnamefont {H.}~\bibnamefont
  {Ritsch}},\ }\href {\doibase 10.1103/PhysRevLett.102.020403} {\bibfield
  {journal} {\bibinfo  {journal} {Phys. Rev. Lett.}\ }\textbf {\bibinfo
  {volume} {102}},\ \bibinfo {pages} {020403} (\bibinfo {year}
  {2009})}\BibitemShut {NoStop}%
\bibitem [{\citenamefont {Wade}\ \emph {et~al.}(2015)\citenamefont {Wade},
  \citenamefont {Sherson},\ and\ \citenamefont
  {M\o{}lmer}}]{PhysRevLett.115.060401}%
  \BibitemOpen
  \bibfield  {author} {\bibinfo {author} {\bibfnamefont {A.~C.~J.}\
  \bibnamefont {Wade}}, \bibinfo {author} {\bibfnamefont {J.~F.}\ \bibnamefont
  {Sherson}}, \ and\ \bibinfo {author} {\bibfnamefont {K.}~\bibnamefont
  {M\o{}lmer}},\ }\href {\doibase 10.1103/PhysRevLett.115.060401} {\bibfield
  {journal} {\bibinfo  {journal} {Phys. Rev. Lett.}\ }\textbf {\bibinfo
  {volume} {115}},\ \bibinfo {pages} {060401} (\bibinfo {year}
  {2015})}\BibitemShut {NoStop}%
\bibitem [{\citenamefont {Ashida}\ and\ \citenamefont
  {Ueda}(2015)}]{PhysRevLett.115.095301}%
  \BibitemOpen
  \bibfield  {author} {\bibinfo {author} {\bibfnamefont {Y.}~\bibnamefont
  {Ashida}}\ and\ \bibinfo {author} {\bibfnamefont {M.}~\bibnamefont {Ueda}},\
  }\href {\doibase 10.1103/PhysRevLett.115.095301} {\bibfield  {journal}
  {\bibinfo  {journal} {Phys. Rev. Lett.}\ }\textbf {\bibinfo {volume} {115}},\
  \bibinfo {pages} {095301} (\bibinfo {year} {2015})}\BibitemShut {NoStop}%
\bibitem [{\citenamefont {Yang}\ \emph {et~al.}(2017)\citenamefont {Yang},
  \citenamefont {Laflamme}, \citenamefont {Vasilyev}, \citenamefont {Baranov},\
  and\ \citenamefont {Zoller}}]{Yang:2017aa}%
  \BibitemOpen
  \bibfield  {author} {\bibinfo {author} {\bibfnamefont {D.}~\bibnamefont
  {Yang}}, \bibinfo {author} {\bibfnamefont {C.}~\bibnamefont {Laflamme}},
  \bibinfo {author} {\bibfnamefont {D.~V.}\ \bibnamefont {Vasilyev}}, \bibinfo
  {author} {\bibfnamefont {M.~A.}\ \bibnamefont {Baranov}}, \ and\ \bibinfo
  {author} {\bibfnamefont {P.}~\bibnamefont {Zoller}},\ }\href{\doibase 10.1103/PhysRevA.98.023852} 
{\bibfield {journal} {\bibinfo {journal} {Phys. Rev. A}\ }\textbf {\bibinfo {volume} {98}},\ 
  \bibinfo {pages} {023852} (\bibinfo {year} {2018})}\BibitemShut {NoStop}%
\bibitem [{\citenamefont {Ritsch}\ \emph {et~al.}(2013)\citenamefont {Ritsch},
  \citenamefont {Domokos}, \citenamefont {Brennecke},\ and\ \citenamefont
  {Esslinger}}]{Ritsch:2013aa}%
  \BibitemOpen
  \bibfield  {author} {\bibinfo {author} {\bibfnamefont {H.}~\bibnamefont
  {Ritsch}}, \bibinfo {author} {\bibfnamefont {P.}~\bibnamefont {Domokos}},
  \bibinfo {author} {\bibfnamefont {F.}~\bibnamefont {Brennecke}}, \ and\
  \bibinfo {author} {\bibfnamefont {T.}~\bibnamefont {Esslinger}},\ }\href
  {\doibase 10.1103/RevModPhys.85.553} {\bibfield  {journal} {\bibinfo
  {journal} {Rev. Mod. Phys.}\ }\textbf {\bibinfo {volume} {85}},\ \bibinfo
  {pages} {553} (\bibinfo {year} {2013})}\BibitemShut {NoStop}%
\bibitem [{\citenamefont {Gardiner}\ and\ \citenamefont
  {Zoller}(2004)}]{gardiner2004}%
  \BibitemOpen
  \bibfield  {author} {\bibinfo {author} {\bibfnamefont {C.}~\bibnamefont
  {Gardiner}}\ and\ \bibinfo {author} {\bibfnamefont {P.}~\bibnamefont
  {Zoller}},\ }\href@noop {} {\emph {\bibinfo {title} {Quantum noise: a
  handbook of Markovian and non-Markovian quantum stochastic methods with
  applications to quantum optics}}},\ Vol.~\bibinfo {volume} {56}\ (\bibinfo
  {publisher} {Springer Science \& Business Media},\ \bibinfo {year}
  {2004})\BibitemShut {NoStop}%
\bibitem [{\citenamefont {Colombe}\ \emph {et~al.}(2007)\citenamefont
  {Colombe}, \citenamefont {Steinmetz}, \citenamefont {Dubois}, \citenamefont
  {Linke}, \citenamefont {Hunger},\ and\ \citenamefont
  {Reichel}}]{2007Natur.450..272C}%
  \BibitemOpen
\bibfield  {journal} {  }\bibfield  {author} {\bibinfo {author} {\bibfnamefont
  {Y.}~\bibnamefont {Colombe}}, \bibinfo {author} {\bibfnamefont
  {T.}~\bibnamefont {Steinmetz}}, \bibinfo {author} {\bibfnamefont
  {G.}~\bibnamefont {Dubois}}, \bibinfo {author} {\bibfnamefont
  {F.}~\bibnamefont {Linke}}, \bibinfo {author} {\bibfnamefont
  {D.}~\bibnamefont {Hunger}}, \ and\ \bibinfo {author} {\bibfnamefont
  {J.}~\bibnamefont {Reichel}},\ }\href@noop {} {\bibfield  {journal} {\bibinfo
   {journal} {Nature}\ }\textbf {\bibinfo {volume} {450}},\ \bibinfo {pages}
  {272} (\bibinfo {year} {2007})}\BibitemShut {NoStop}%
\bibitem [{\citenamefont {Brennecke}\ \emph {et~al.}(2007)\citenamefont
  {Brennecke}, \citenamefont {Donner}, \citenamefont {Ritter}, \citenamefont
  {Bourdel}, \citenamefont {K{\"o}hl},\ and\ \citenamefont
  {Esslinger}}]{Brennecke:2007aa}%
  \BibitemOpen
  \bibfield  {author} {\bibinfo {author} {\bibfnamefont {F.}~\bibnamefont
  {Brennecke}}, \bibinfo {author} {\bibfnamefont {T.}~\bibnamefont {Donner}},
  \bibinfo {author} {\bibfnamefont {S.}~\bibnamefont {Ritter}}, \bibinfo
  {author} {\bibfnamefont {T.}~\bibnamefont {Bourdel}}, \bibinfo {author}
  {\bibfnamefont {M.}~\bibnamefont {K{\"o}hl}}, \ and\ \bibinfo {author}
  {\bibfnamefont {T.}~\bibnamefont {Esslinger}},\ }\href@noop {} {\bibfield
  {journal} {\bibinfo  {journal} {Nature}\ }\textbf {\bibinfo {volume} {450}},\
  \bibinfo {pages} {268} (\bibinfo {year} {2007})}\BibitemShut {NoStop}%
\bibitem [{\citenamefont {Slama}\ \emph {et~al.}(2007)\citenamefont {Slama},
  \citenamefont {Krenz}, \citenamefont {Bux}, \citenamefont {Zimmermann},\ and\
  \citenamefont {Courteille}}]{PhysRevA.75.063620}%
  \BibitemOpen
  \bibfield  {author} {\bibinfo {author} {\bibfnamefont {S.}~\bibnamefont
  {Slama}}, \bibinfo {author} {\bibfnamefont {G.}~\bibnamefont {Krenz}},
  \bibinfo {author} {\bibfnamefont {S.}~\bibnamefont {Bux}}, \bibinfo {author}
  {\bibfnamefont {C.}~\bibnamefont {Zimmermann}}, \ and\ \bibinfo {author}
  {\bibfnamefont {P.~W.}\ \bibnamefont {Courteille}},\ }\href {\doibase
  10.1103/PhysRevA.75.063620} {\bibfield  {journal} {\bibinfo  {journal} {Phys.
  Rev. A}\ }\textbf {\bibinfo {volume} {75}},\ \bibinfo {pages} {063620}
  (\bibinfo {year} {2007})}\BibitemShut {NoStop}%
\bibitem [{\citenamefont {Murch}\ \emph {et~al.}(2008)\citenamefont {Murch},
  \citenamefont {Moore}, \citenamefont {Gupta},\ and\ \citenamefont
  {Stamper-Kurn}}]{Murch:2008aa}%
  \BibitemOpen
  \bibfield  {author} {\bibinfo {author} {\bibfnamefont {K.~W.}\ \bibnamefont
  {Murch}}, \bibinfo {author} {\bibfnamefont {K.~L.}\ \bibnamefont {Moore}},
  \bibinfo {author} {\bibfnamefont {S.}~\bibnamefont {Gupta}}, \ and\ \bibinfo
  {author} {\bibfnamefont {D.~M.}\ \bibnamefont {Stamper-Kurn}},\ }\href@noop
  {} {\bibfield  {journal} {\bibinfo  {journal} {Nature Physics}\ }\textbf
  {\bibinfo {volume} {4}},\ \bibinfo {pages} {561} (\bibinfo {year}
  {2008})}\BibitemShut {NoStop}%
\bibitem [{\citenamefont {Kessler}\ \emph {et~al.}(2014)\citenamefont
  {Kessler}, \citenamefont {Klinder}, \citenamefont {Wolke},\ and\
  \citenamefont {Hemmerich}}]{Keler:2014aa}%
  \BibitemOpen
  \bibfield  {author} {\bibinfo {author} {\bibfnamefont {H.}~\bibnamefont
  {Kessler}}, \bibinfo {author} {\bibfnamefont {J.}~\bibnamefont {Klinder}},
  \bibinfo {author} {\bibfnamefont {M.}~\bibnamefont {Wolke}}, \ and\ \bibinfo
  {author} {\bibfnamefont {A.}~\bibnamefont {Hemmerich}},\ }\href
  {http://stacks.iop.org/1367-2630/16/i=5/a=053008} {\bibfield  {journal}
  {\bibinfo  {journal} {New Journal of Physics}\ }\textbf {\bibinfo {volume}
  {16}},\ \bibinfo {pages} {053008} (\bibinfo {year} {2014})}\BibitemShut
  {NoStop}%
\bibitem [{\citenamefont {Braaten}\ and\ \citenamefont
  {Platter}(2008)}]{PhysRevLett.100.205301}%
  \BibitemOpen
  \bibfield  {author} {\bibinfo {author} {\bibfnamefont {E.}~\bibnamefont
  {Braaten}}\ and\ \bibinfo {author} {\bibfnamefont {L.}~\bibnamefont
  {Platter}},\ }\href {\doibase 10.1103/PhysRevLett.100.205301} {\bibfield
  {journal} {\bibinfo  {journal} {Phys. Rev. Lett.}\ }\textbf {\bibinfo
  {volume} {100}},\ \bibinfo {pages} {205301} (\bibinfo {year}
  {2008})}\BibitemShut {NoStop}%
\bibitem [{\citenamefont {Son}\ and\ \citenamefont
  {Thompson}(2010)}]{PhysRevA.81.063634}%
  \BibitemOpen
  \bibfield  {author} {\bibinfo {author} {\bibfnamefont {D.~T.}\ \bibnamefont
  {Son}}\ and\ \bibinfo {author} {\bibfnamefont {E.~G.}\ \bibnamefont
  {Thompson}},\ }\href {\doibase 10.1103/PhysRevA.81.063634} {\bibfield
  {journal} {\bibinfo  {journal} {Phys. Rev. A}\ }\textbf {\bibinfo {volume}
  {81}},\ \bibinfo {pages} {063634} (\bibinfo {year} {2010})}\BibitemShut
  {NoStop}%
\bibitem [{\citenamefont {Hofmann}(2011)}]{PhysRevA.84.043603}%
  \BibitemOpen
  \bibfield  {author} {\bibinfo {author} {\bibfnamefont {J.}~\bibnamefont
  {Hofmann}},\ }\href {\doibase 10.1103/PhysRevA.84.043603} {\bibfield
  {journal} {\bibinfo  {journal} {Phys. Rev. A}\ }\textbf {\bibinfo {volume}
  {84}},\ \bibinfo {pages} {043603} (\bibinfo {year} {2011})}\BibitemShut
  {NoStop}%
\bibitem [{\citenamefont {Goldberger}\ and\ \citenamefont
  {Rothstein}(2012)}]{PhysRevA.85.013613}%
  \BibitemOpen
  \bibfield  {author} {\bibinfo {author} {\bibfnamefont {W.~D.}\ \bibnamefont
  {Goldberger}}\ and\ \bibinfo {author} {\bibfnamefont {I.~Z.}\ \bibnamefont
  {Rothstein}},\ }\href {\doibase 10.1103/PhysRevA.85.013613} {\bibfield
  {journal} {\bibinfo  {journal} {Phys. Rev. A}\ }\textbf {\bibinfo {volume}
  {85}},\ \bibinfo {pages} {013613} (\bibinfo {year} {2012})}\BibitemShut
  {NoStop}%
\bibitem [{\citenamefont {Hofmann}\ and\ \citenamefont
  {Zwerger}(2017)}]{PhysRevX.7.011022}%
  \BibitemOpen
  \bibfield  {author} {\bibinfo {author} {\bibfnamefont {J.}~\bibnamefont
  {Hofmann}}\ and\ \bibinfo {author} {\bibfnamefont {W.}~\bibnamefont
  {Zwerger}},\ }\href {\doibase 10.1103/PhysRevX.7.011022} {\bibfield
  {journal} {\bibinfo  {journal} {Phys. Rev. X}\ }\textbf {\bibinfo {volume}
  {7}},\ \bibinfo {pages} {011022} (\bibinfo {year} {2017})}\BibitemShut
  {NoStop}%
\bibitem [{\citenamefont {Hoinka}\ \emph {et~al.}(2012)\citenamefont {Hoinka},
  \citenamefont {Lingham}, \citenamefont {Delehaye},\ and\ \citenamefont
  {Vale}}]{PhysRevLett.109.050403}%
  \BibitemOpen
  \bibfield  {author} {\bibinfo {author} {\bibfnamefont {S.}~\bibnamefont
  {Hoinka}}, \bibinfo {author} {\bibfnamefont {M.}~\bibnamefont {Lingham}},
  \bibinfo {author} {\bibfnamefont {M.}~\bibnamefont {Delehaye}}, \ and\
  \bibinfo {author} {\bibfnamefont {C.~J.}\ \bibnamefont {Vale}},\ }\href
  {\doibase 10.1103/PhysRevLett.109.050403} {\bibfield  {journal} {\bibinfo
  {journal} {Phys. Rev. Lett.}\ }\textbf {\bibinfo {volume} {109}},\ \bibinfo
  {pages} {050403} (\bibinfo {year} {2012})}\BibitemShut {NoStop}%
\bibitem [{\citenamefont {Hoinka}\ \emph {et~al.}(2013)\citenamefont {Hoinka},
  \citenamefont {Lingham}, \citenamefont {Fenech}, \citenamefont {Hu},
  \citenamefont {Vale}, \citenamefont {Drut},\ and\ \citenamefont
  {Gandolfi}}]{Hoinka:2013aa}%
  \BibitemOpen
  \bibfield  {author} {\bibinfo {author} {\bibfnamefont {S.}~\bibnamefont
  {Hoinka}}, \bibinfo {author} {\bibfnamefont {M.}~\bibnamefont {Lingham}},
  \bibinfo {author} {\bibfnamefont {K.}~\bibnamefont {Fenech}}, \bibinfo
  {author} {\bibfnamefont {H.}~\bibnamefont {Hu}}, \bibinfo {author}
  {\bibfnamefont {C.~J.}\ \bibnamefont {Vale}}, \bibinfo {author}
  {\bibfnamefont {J.~E.}\ \bibnamefont {Drut}}, \ and\ \bibinfo {author}
  {\bibfnamefont {S.}~\bibnamefont {Gandolfi}},\ }\href {\doibase
  10.1103/PhysRevLett.110.055305} {\bibfield  {journal} {\bibinfo  {journal}
  {Phys. Rev. Lett.}\ }\textbf {\bibinfo {volume} {110}},\ \bibinfo {pages}
  {055305} (\bibinfo {year} {2013})}\BibitemShut {NoStop}%
\bibitem [{\citenamefont {Shimizu}(1991)}]{PhysRevA.43.3819}%
  \BibitemOpen
  \bibfield  {author} {\bibinfo {author} {\bibfnamefont {A.}~\bibnamefont
  {Shimizu}},\ }\href {\doibase 10.1103/PhysRevA.43.3819} {\bibfield  {journal}
  {\bibinfo  {journal} {Phys. Rev. A}\ }\textbf {\bibinfo {volume} {43}},\
  \bibinfo {pages} {3819} (\bibinfo {year} {1991})}\BibitemShut {NoStop}%
\bibitem [{\citenamefont {Yang}\ \emph {et~al.}(2018)\citenamefont {Yang},
  \citenamefont {Laflamme}, \citenamefont {Vasilyev}, \citenamefont {Baranov},\
  and\ \citenamefont {Zoller}}]{PhysRevLett.120.133601}%
  \BibitemOpen
  \bibfield  {author} {\bibinfo {author} {\bibfnamefont {D.}~\bibnamefont
  {Yang}}, \bibinfo {author} {\bibfnamefont {C.}~\bibnamefont {Laflamme}},
  \bibinfo {author} {\bibfnamefont {D.~V.}\ \bibnamefont {Vasilyev}}, \bibinfo
  {author} {\bibfnamefont {M.~A.}\ \bibnamefont {Baranov}}, \ and\ \bibinfo
  {author} {\bibfnamefont {P.}~\bibnamefont {Zoller}},\ }\href {\doibase
  10.1103/PhysRevLett.120.133601} {\bibfield  {journal} {\bibinfo  {journal}
  {Phys. Rev. Lett.}\ }\textbf {\bibinfo {volume} {120}},\ \bibinfo {pages}
  {133601} (\bibinfo {year} {2018})}\BibitemShut {NoStop}%
 \bibitem [{\citenamefont {Gerbier}\ and\ \citenamefont
  {Castin}(2010)}]{Gerbier:2010aa}%
  \BibitemOpen
  \bibfield  {author} {\bibinfo {author} {\bibfnamefont {F.}~\bibnamefont
  {Gerbier}}\ and\ \bibinfo {author} {\bibfnamefont {Y.}~\bibnamefont
  {Castin}},\ }\href {\doibase 10.1103/PhysRevA.82.013615} {\bibfield
  {journal} {\bibinfo  {journal} {Phys. Rev. A}\ }\textbf {\bibinfo {volume}
  {82}},\ \bibinfo {pages} {013615} (\bibinfo {year} {2010})}\BibitemShut
  {NoStop}%
\bibitem [{\citenamefont {Landig}\ \emph {et~al.}(2015)\citenamefont {Landig},
  \citenamefont {Brennecke}, \citenamefont {Mottl}, \citenamefont {Donner},\
  and\ \citenamefont {Esslinger}}]{landig2015aa}%
  \BibitemOpen
  \bibfield  {author} {\bibinfo {author} {\bibfnamefont {R.}~\bibnamefont
  {Landig}}, \bibinfo {author} {\bibfnamefont {F.}~\bibnamefont {Brennecke}},
  \bibinfo {author} {\bibfnamefont {R.}~\bibnamefont {Mottl}}, \bibinfo
  {author} {\bibfnamefont {T.}~\bibnamefont {Donner}}, \ and\ \bibinfo {author}
  {\bibfnamefont {T.}~\bibnamefont {Esslinger}},\ }\href@noop {} {\bibfield
  {journal} {\bibinfo  {journal} {Nature communications}\ }\textbf {\bibinfo
  {volume} {6}},\ \bibinfo {pages} {7046} (\bibinfo {year} {2015})}\BibitemShut
  {NoStop}%
\bibitem [{\citenamefont {Larson}\ \emph {et~al.}(2008)\citenamefont {Larson},
  \citenamefont {Morigi},\ and\ \citenamefont
  {Lewenstein}}]{PhysRevA.78.023815}%
  \BibitemOpen
  \bibfield  {author} {\bibinfo {author} {\bibfnamefont {J.}~\bibnamefont
  {Larson}}, \bibinfo {author} {\bibfnamefont {G.}~\bibnamefont {Morigi}}, \
  and\ \bibinfo {author} {\bibfnamefont {M.}~\bibnamefont {Lewenstein}},\
  }\href {\doibase 10.1103/PhysRevA.78.023815} {\bibfield  {journal} {\bibinfo
  {journal} {Phys. Rev. A}\ }\textbf {\bibinfo {volume} {78}},\ \bibinfo
  {pages} {023815} (\bibinfo {year} {2008})}\BibitemShut {NoStop}%
\bibitem [{\citenamefont {Kanamoto}\ and\ \citenamefont
  {Meystre}(2010)}]{PhysRevLett.104.063601}%
  \BibitemOpen
  \bibfield  {author} {\bibinfo {author} {\bibfnamefont {R.}~\bibnamefont
  {Kanamoto}}\ and\ \bibinfo {author} {\bibfnamefont {P.}~\bibnamefont
  {Meystre}},\ }\href {\doibase 10.1103/PhysRevLett.104.063601} {\bibfield
  {journal} {\bibinfo  {journal} {Phys. Rev. Lett.}\ }\textbf {\bibinfo
  {volume} {104}},\ \bibinfo {pages} {063601} (\bibinfo {year}
  {2010})}\BibitemShut {NoStop}%
\bibitem [{\citenamefont {Sun}\ \emph {et~al.}(2011)\citenamefont {Sun},
  \citenamefont {Hu}, \citenamefont {Ji},\ and\ \citenamefont
  {Liu}}]{PhysRevA.83.043606}%
  \BibitemOpen
  \bibfield  {author} {\bibinfo {author} {\bibfnamefont {Q.}~\bibnamefont
  {Sun}}, \bibinfo {author} {\bibfnamefont {X.-H.}\ \bibnamefont {Hu}},
  \bibinfo {author} {\bibfnamefont {A.-C.}\ \bibnamefont {Ji}}, \ and\ \bibinfo
  {author} {\bibfnamefont {W.~M.}\ \bibnamefont {Liu}},\ }\href {\doibase
  10.1103/PhysRevA.83.043606} {\bibfield  {journal} {\bibinfo  {journal} {Phys.
  Rev. A}\ }\textbf {\bibinfo {volume} {83}},\ \bibinfo {pages} {043606}
  (\bibinfo {year} {2011})}\BibitemShut {NoStop}%
\bibitem [{\citenamefont {Ingold}\ and\ \citenamefont
  {Nazarov}(1992)}]{Ingold:1992aa}%
  \BibitemOpen
  \bibfield  {author} {\bibinfo {author} {\bibfnamefont {G.-L.}\ \bibnamefont
  {Ingold}}\ and\ \bibinfo {author} {\bibfnamefont {Y.~V.}\ \bibnamefont
  {Nazarov}},\ }\enquote {\bibinfo {title} {Charge tunneling rates in
  ultrasmall junctions},}\ in\ \href {\doibase 10.1007/978-1-4757-2166-9{\_}2}
  {\emph {\bibinfo {booktitle} {Single Charge Tunneling: Coulomb Blockade
  Phenomena In Nanostructures}}},\ \bibinfo {editor} {edited by\ \bibinfo
  {editor} {\bibfnamefont {H.}~\bibnamefont {Grabert}}\ and\ \bibinfo {editor}
  {\bibfnamefont {M.~H.}\ \bibnamefont {Devoret}}}\ (\bibinfo  {publisher}
  {Springer US},\ \bibinfo {address} {Boston, MA},\ \bibinfo {year} {1992})\
  pp.\ \bibinfo {pages} {21--107}\BibitemShut {NoStop}%
\bibitem [{\citenamefont {Altimiras}\ \emph {et~al.}(2016)\citenamefont
  {Altimiras}, \citenamefont {Portier},\ and\ \citenamefont
  {Joyez}}]{PhysRevX.6.031002}%
  \BibitemOpen
  \bibfield  {author} {\bibinfo {author} {\bibfnamefont {C.}~\bibnamefont
  {Altimiras}}, \bibinfo {author} {\bibfnamefont {F.}~\bibnamefont {Portier}},
  \ and\ \bibinfo {author} {\bibfnamefont {P.}~\bibnamefont {Joyez}},\ }\href
  {\doibase 10.1103/PhysRevX.6.031002} {\bibfield  {journal} {\bibinfo
  {journal} {Phys. Rev. X}\ }\textbf {\bibinfo {volume} {6}},\ \bibinfo {pages}
  {031002} (\bibinfo {year} {2016})}\BibitemShut {NoStop}%
\bibitem [{\citenamefont {Clerk}\ \emph {et~al.}(2010)\citenamefont {Clerk},
  \citenamefont {Devoret}, \citenamefont {Girvin}, \citenamefont {Marquardt},\
  and\ \citenamefont {Schoelkopf}}]{Clerk:2010aa}%
  \BibitemOpen
  \bibfield  {author} {\bibinfo {author} {\bibfnamefont {A.~A.}\ \bibnamefont
  {Clerk}}, \bibinfo {author} {\bibfnamefont {M.~H.}\ \bibnamefont {Devoret}},
  \bibinfo {author} {\bibfnamefont {S.~M.}\ \bibnamefont {Girvin}}, \bibinfo
  {author} {\bibfnamefont {F.}~\bibnamefont {Marquardt}}, \ and\ \bibinfo
  {author} {\bibfnamefont {R.~J.}\ \bibnamefont {Schoelkopf}},\ }\href
  {\doibase 10.1103/RevModPhys.82.1155} {\bibfield  {journal} {\bibinfo
  {journal} {Rev. Mod. Phys.}\ }\textbf {\bibinfo {volume} {82}},\ \bibinfo
  {pages} {1155} (\bibinfo {year} {2010})}\BibitemShut {NoStop}%
\bibitem [{\citenamefont {Aspelmeyer}\ \emph {et~al.}(2014)\citenamefont
  {Aspelmeyer}, \citenamefont {Kippenberg},\ and\ \citenamefont
  {Marquardt}}]{Aspelmeyer:2014aa}%
  \BibitemOpen
  \bibfield  {author} {\bibinfo {author} {\bibfnamefont {M.}~\bibnamefont
  {Aspelmeyer}}, \bibinfo {author} {\bibfnamefont {T.~J.}\ \bibnamefont
  {Kippenberg}}, \ and\ \bibinfo {author} {\bibfnamefont {F.}~\bibnamefont
  {Marquardt}},\ }\href {\doibase 10.1103/RevModPhys.86.1391} {\bibfield
  {journal} {\bibinfo  {journal} {Rev. Mod. Phys.}\ }\textbf {\bibinfo {volume}
  {86}},\ \bibinfo {pages} {1391} (\bibinfo {year} {2014})}\BibitemShut
  {NoStop}%
\bibitem [{\citenamefont {H\"ausler}\ \emph {et~al.}(2017)\citenamefont
  {H\"ausler}, \citenamefont {Nakajima}, \citenamefont {Lebrat}, \citenamefont
  {Husmann}, \citenamefont {Krinner}, \citenamefont {Esslinger},\ and\
  \citenamefont {Brantut}}]{PhysRevLett.119.030403}%
  \BibitemOpen
  \bibfield  {author} {\bibinfo {author} {\bibfnamefont {S.}~\bibnamefont
  {H\"ausler}}, \bibinfo {author} {\bibfnamefont {S.}~\bibnamefont {Nakajima}},
  \bibinfo {author} {\bibfnamefont {M.}~\bibnamefont {Lebrat}}, \bibinfo
  {author} {\bibfnamefont {D.}~\bibnamefont {Husmann}}, \bibinfo {author}
  {\bibfnamefont {S.}~\bibnamefont {Krinner}}, \bibinfo {author} {\bibfnamefont
  {T.}~\bibnamefont {Esslinger}}, \ and\ \bibinfo {author} {\bibfnamefont
  {J.-P.}\ \bibnamefont {Brantut}},\ }\href {\doibase
  10.1103/PhysRevLett.119.030403} {\bibfield  {journal} {\bibinfo  {journal}
  {Phys. Rev. Lett.}\ }\textbf {\bibinfo {volume} {119}},\ \bibinfo {pages}
  {030403} (\bibinfo {year} {2017})}\BibitemShut {NoStop}%
\bibitem [{\citenamefont {Kollath}\ \emph {et~al.}(2016)\citenamefont
  {Kollath}, \citenamefont {Sheikhan}, \citenamefont {Wolff},\ and\
  \citenamefont {Brennecke}}]{PhysRevLett.116.060401}%
  \BibitemOpen
  \bibfield  {author} {\bibinfo {author} {\bibfnamefont {C.}~\bibnamefont
  {Kollath}}, \bibinfo {author} {\bibfnamefont {A.}~\bibnamefont {Sheikhan}},
  \bibinfo {author} {\bibfnamefont {S.}~\bibnamefont {Wolff}}, \ and\ \bibinfo
  {author} {\bibfnamefont {F.}~\bibnamefont {Brennecke}},\ }\href {\doibase
  10.1103/PhysRevLett.116.060401} {\bibfield  {journal} {\bibinfo  {journal}
  {Phys. Rev. Lett.}\ }\textbf {\bibinfo {volume} {116}},\ \bibinfo {pages}
  {060401} (\bibinfo {year} {2016})}\BibitemShut {NoStop}%
\bibitem [{\citenamefont {Zheng}\ and\ \citenamefont
  {Cooper}(2016)}]{PhysRevLett.117.175302}%
  \BibitemOpen
  \bibfield  {author} {\bibinfo {author} {\bibfnamefont {W.}~\bibnamefont
  {Zheng}}\ and\ \bibinfo {author} {\bibfnamefont {N.~R.}\ \bibnamefont
  {Cooper}},\ }\href {\doibase 10.1103/PhysRevLett.117.175302} {\bibfield
  {journal} {\bibinfo  {journal} {Phys. Rev. Lett.}\ }\textbf {\bibinfo
  {volume} {117}},\ \bibinfo {pages} {175302} (\bibinfo {year}
  {2016})}\BibitemShut {NoStop}%
\bibitem [{\citenamefont {Laflamme}\ \emph {et~al.}(2017)\citenamefont
  {Laflamme}, \citenamefont {Yang},\ and\ \citenamefont
  {Zoller}}]{PhysRevA.95.043843}%
  \BibitemOpen
  \bibfield  {author} {\bibinfo {author} {\bibfnamefont {C.}~\bibnamefont
  {Laflamme}}, \bibinfo {author} {\bibfnamefont {D.}~\bibnamefont {Yang}}, \
  and\ \bibinfo {author} {\bibfnamefont {P.}~\bibnamefont {Zoller}},\ }\href
  {\doibase 10.1103/PhysRevA.95.043843} {\bibfield  {journal} {\bibinfo
  {journal} {Phys. Rev. A}\ }\textbf {\bibinfo {volume} {95}},\ \bibinfo
  {pages} {043843} (\bibinfo {year} {2017})}\BibitemShut {NoStop}%
\bibitem [{\citenamefont {Brantut}\ \emph {et~al.}(2013)\citenamefont
  {Brantut}, \citenamefont {Grenier}, \citenamefont {Meineke}, \citenamefont
  {Stadler}, \citenamefont {Krinner}, \citenamefont {Kollath}, \citenamefont
  {Esslinger},\ and\ \citenamefont {Georges}}]{Brantut:2013aa}%
  \BibitemOpen
  \bibfield  {author} {\bibinfo {author} {\bibfnamefont {J.-P.}\ \bibnamefont
  {Brantut}}, \bibinfo {author} {\bibfnamefont {C.}~\bibnamefont {Grenier}},
  \bibinfo {author} {\bibfnamefont {J.}~\bibnamefont {Meineke}}, \bibinfo
  {author} {\bibfnamefont {D.}~\bibnamefont {Stadler}}, \bibinfo {author}
  {\bibfnamefont {S.}~\bibnamefont {Krinner}}, \bibinfo {author} {\bibfnamefont
  {C.}~\bibnamefont {Kollath}}, \bibinfo {author} {\bibfnamefont
  {T.}~\bibnamefont {Esslinger}}, \ and\ \bibinfo {author} {\bibfnamefont
  {A.}~\bibnamefont {Georges}},\ }\href
  {http://www.sciencemag.org/content/342/6159/713.abstract} {\bibfield
  {journal} {\bibinfo  {journal} {Science}\ }\textbf {\bibinfo {volume}
  {342}},\ \bibinfo {pages} {713} (\bibinfo {year} {2013})}\BibitemShut
  {NoStop}%
\bibitem [{\citenamefont {Krinner}\ \emph {et~al.}(2016)\citenamefont
  {Krinner}, \citenamefont {Lebrat}, \citenamefont {Husmann}, \citenamefont
  {Grenier}, \citenamefont {Brantut},\ and\ \citenamefont
  {Esslinger}}]{Krinner:2016ab}%
  \BibitemOpen
  \bibfield  {author} {\bibinfo {author} {\bibfnamefont {S.}~\bibnamefont
  {Krinner}}, \bibinfo {author} {\bibfnamefont {M.}~\bibnamefont {Lebrat}},
  \bibinfo {author} {\bibfnamefont {D.}~\bibnamefont {Husmann}}, \bibinfo
  {author} {\bibfnamefont {C.}~\bibnamefont {Grenier}}, \bibinfo {author}
  {\bibfnamefont {J.-P.}\ \bibnamefont {Brantut}}, \ and\ \bibinfo {author}
  {\bibfnamefont {T.}~\bibnamefont {Esslinger}},\ }\href
  {http://www.pnas.org/content/113/29/8144.abstract N2 - We study particle and
  spin transport in a single-mode quantum point contact, using a charge
  neutral, quantum degenerate Fermi gas with tunable, attractive interactions.
  This yields the spin and particle conductance of the point contact as a
  function of chemical potential or confinement. The measurements cover a
  regime from weak attraction, where quantized conductance is observed, to the
  resonantly interacting superfluid. Spin conductance exhibits a broad maximum
  when varying the chemical potential at moderate interactions, which signals
  the emergence of Cooper pairing. In contrast, the particle conductance is
  unexpectedly enhanced even before the gas is expected to turn into a
  superfluid, continuously rising from the plateau at 1/h for weak interactions
  to plateau-like features at nonuniversal values as high as 4/h for
  intermediate interactions. For strong interactions, the particle conductance
  plateaus disappear and the spin conductance gets suppressed, confirming the
  spin-insulating character of a superfluid. Our observations document the
  breakdown of universal conductance quantization as many-body correlations
  appear. The observed anomalous quantization challenges a Fermi liquid
  description of the normal phase, shedding new light on the nature of the
  strongly attractive Fermi gas.} {\bibfield  {journal} {\bibinfo  {journal}
  {Proceedings of the National Academy of Sciences}\ }\textbf {\bibinfo
  {volume} {113}},\ \bibinfo {pages} {8144} (\bibinfo {year}
  {2016})}\BibitemShut {NoStop}%
\bibitem [{\citenamefont {Cox}\ \emph {et~al.}(2016)\citenamefont {Cox},
  \citenamefont {Greve}, \citenamefont {Wu},\ and\ \citenamefont
  {Thompson}}]{PhysRevA.94.061601}%
  \BibitemOpen
  \bibfield  {author} {\bibinfo {author} {\bibfnamefont {K.~C.}\ \bibnamefont
  {Cox}}, \bibinfo {author} {\bibfnamefont {G.~P.}\ \bibnamefont {Greve}},
  \bibinfo {author} {\bibfnamefont {B.}~\bibnamefont {Wu}}, \ and\ \bibinfo
  {author} {\bibfnamefont {J.~K.}\ \bibnamefont {Thompson}},\ }\href {\doibase
  10.1103/PhysRevA.94.061601} {\bibfield  {journal} {\bibinfo  {journal} {Phys.
  Rev. A}\ }\textbf {\bibinfo {volume} {94}},\ \bibinfo {pages} {061601}
  (\bibinfo {year} {2016})}\BibitemShut {NoStop}%
\bibitem [{\citenamefont {Vallet}\ \emph {et~al.}(2017)\citenamefont {Vallet},
  \citenamefont {Bookjans}, \citenamefont {Eismann}, \citenamefont {Bilicki},
  \citenamefont {Targat},\ and\ \citenamefont
  {Lodewyck}}]{1367-2630-19-8-083002}%
  \BibitemOpen
  \bibfield  {author} {\bibinfo {author} {\bibfnamefont {G.}~\bibnamefont
  {Vallet}}, \bibinfo {author} {\bibfnamefont {E.}~\bibnamefont {Bookjans}},
  \bibinfo {author} {\bibfnamefont {U.}~\bibnamefont {Eismann}}, \bibinfo
  {author} {\bibfnamefont {S.}~\bibnamefont {Bilicki}}, \bibinfo {author}
  {\bibfnamefont {R.~L.}\ \bibnamefont {Targat}}, \ and\ \bibinfo {author}
  {\bibfnamefont {J.}~\bibnamefont {Lodewyck}},\ }\href
  {http://stacks.iop.org/1367-2630/19/i=8/a=083002} {\bibfield  {journal}
  {\bibinfo  {journal} {New Journal of Physics}\ }\textbf {\bibinfo {volume}
  {19}},\ \bibinfo {pages} {083002} (\bibinfo {year} {2017})}\BibitemShut
  {NoStop}%
\end{thebibliography}
%

\end{document}